\documentclass{raa}

\usepackage{graphicx, times}
\usepackage{natbib}
\usepackage{amssymb,amsmath}
\usepackage{bm}

\begin{document}
%Issue 0.06 - revisions for publication
\title{Chemo-dynamical modelling with Schwarzschild's method}

%\volnopage{Vol.0 (201x) No.0, 000--000}      %%preserved for Editor. DOn't remove!
%\setcounter{page}{1}          %%starting page, preserved for Editor. DOn't remove!

\author{R. J. Long\inst{1,2}, Shude Mao\inst{3,1,2}}

\institute{National Astronomical Observatories, Chinese Academy of Sciences, 20A Datun Rd, Chaoyang District, Beijing 100012, China; {\it rjl@nao.cas.cn}\\
        \and
           Jodrell Bank Centre for Astrophysics, School of Physics and Astronomy, The University of Manchester, Oxford Road, Manchester M13 9PL, UK\\
        \and
        	    Physics Department and Tsinghua Centre for Astrophysics, Tsinghua University, Beijing, 100084, China
           }

\date{Received~~2018 month day; accepted~~2018~~month day}

\abstract{
We extend Schwarzschild's dynamical modelling method to model absorption line strength data as well as the more usual luminosity and kinematic data.  Our approach draws on earlier published work by the first author with the Syer \& Tremaine made-to-measure (M2M) dynamical modelling method and uses similar ideas to create a chemo-Schwarzschild method.  We apply our extended Schwarzschild method to the same four early type galaxies (NGC 1248, NGC 3838, NGC 4452, NGC 4551) as the chemo-M2M work, and are able to recover successfully the 2D absorption line strength for the three lines we model (H$\beta$, Fe5015, Mg$\,b$).  We believe that this is the first time Schwarzschild's method has been used in this way. The techniques developed can be applied to modelling other aspects of galaxies, for example age and metallicity data coming from stellar population modelling, and are not specific to absorption line strength data.
}

\authorrunning{R. J. Long \& Shude Mao}            %author_head in even pages
\titlerunning{Chemo-dynamical modelling with Schwarzschild's method}  % title_head in odd pages

\maketitle

%% The author head (on even pages) and the title head (on odd pages) will be
%% automatically extracted from \author{} and \title{}. Whenever the title is too long,
%% you will be asked to supply a shorter one by inserting either \authorrunning{} or
%% \titlerunning{} before \maketitle. Anyway, you can specify your own heads.
%%
%%
%% Note: In the following text body of your manuscript, please note several differences from
%%       other major journals:
%% (1) \subsection{Please Capitalize the First Letter of Each Notional Word in Subsection Title}
%% (2) Please Capitalize the First Letter of Each Notional Word in all tables' captions

\section{Introduction}\label{sec:introduction}
In \citet{Long2016}, it was stated that ``a galaxy's construction and evolution are imprinted in its kinematics and chemistry but require significant analysis to identify the contributing componentry''.  The paper then demonstrated successfully how the made-to-measure (M2M) method proposed by \citet{Syer1996} for modelling stellar dynamical systems could be extended to model spectral absorption line strength data.    Within the paper, it was noted that a similar approach to chemo-dynamical modelling may be possible using \citet{Schwarz1979} modelling.  We are pleased to report that we have now developed such a modelling method by a simple extension to Schwarzschild's method.  This letter draws heavily on \citet{Long2016} and should be considered as an extension to it.  We believe that this is the first time in which line strength data has been employed in Schwarzschild modelling.

Schwarzschild's method pre-dates the M2M method by almost 20 years and its application to galaxies is more extensive and includes those described for the M2M method in \citet{Long2016}.  Now that we have a chemo-dynamical Schwarzschild's method, we have a further tool to examine and analyse data from from IFU-based galaxy surveys (for example, ATLAS$^{\rm{3D}}$, \citealt{AtlasI}; SAMI, \citealt{SAMI2015}; MaNGA, \citealt{Bundy2015}), hopefully adding to our knowledge of the componentry underlying galaxies and how it was assembled.

As might be expected, our objectives in performing the current investigation mirror those in \citet{Long2016} and are
\begin{enumerate}
\item to extend Schwarzschild's method to model absorption line strength data as well as the more usual luminosity and kinematic data, and to create a software implementation of the revised method,
\item to apply the method to the same selection of external galaxies as \citet{Long2016} and confirm that the criteria for a successful Schwarzschild model can be met (orbits weighted and observables reproduced), and
\item to understand the limitations of the extensions and to identify areas for future work. 
\end{enumerate}

The structure of the paper broadly follows the objectives. In Section \ref{sec:schw}, we describe our enhanced Schwarzschild's method.  In Sections \ref{sec:external} and \ref{sec:results}, we apply the method to four galaxies taken from the ATLAS$^{\rm{3D}}$ survey and summarise our results. In Sections \ref{sec:discuss} and \ref{sec:conclusions}, we discuss the results and draw conclusions identifying areas for further investigation.

\section{Schwarzschild's Method}\label{sec:schw}
Schwarzschild's method was first described in \citet{Schwarz1979}.  The method is concerned with weighting orbits in such a way that observations of a stellar system may be reproduced.  Whereas in the M2M method particles weights are adjusted as the particles are being orbited, in Schwarszschild's method the orbit weights are calculated only after orbit creation has been completed. The method has been enhanced and applied by many researchers since 1979 (for example, \citealt{Rix1997}, \citealt{Valluri2004}, \citealt{Cappellari2006}, \citealt{Remco2008}), and is very well described in those and other papers.  \citet{BT2008} contains an overview.  We therefore do not describe the full detail of the method here but only sufficient so that it is clear how line strength data can be modelled.

In Section \ref{sec:theory}, we frame Schwarzschild's method in terms of matrices as a convex optimisation problem, and then in Section \ref{sec:enhtheory}, using the same terminology, describe how line strength data can be handled.  In essence, for every orbit, we introduce an additional value per orbit for each absorption line to be modelled.

\subsection{Basic Theory - Orbit Weights}\label{sec:theory}
Calculation of the orbit weights in Schwarzschild's method is achieved using a `least squares' approach to minimise the residuals between the model observables and the measured observables of a stellar system.  Given that linear least squares is just a subset of convex function optimisation theory (\citealt{CVX}), we choose to think of the minimisation in those terms.  In matrix form we minimise
\begin{equation}
	\|\mathbf{Dw} - \mathbf{K}\|^2_2,
	\label{eqn:orbwts}
\end{equation}
where $\mathbf{D}$ is the `design' matrix giving individual orbit contributions to model observables, $\mathbf{w}$ represents the orbit weights to be determined, and $\mathbf{K}$ contains the `measured' observables.  The L2-norm (Euclidean norm) is indicated by $\|.....\|_2$.  For our purposes, $\mathbf{K}$ is taken to contain both kinematic and luminosity measurements.  Whether or not luminosity measurements should be taken as range constraints within the minimisation (for example, \citealt{Remco2008}, \citealt{Valluri2004}) does not affect our arguments.  Given that individual orbit weights should not be negative, the least squares algorithm required must be able to generate a non-negative least squares result (see \citealt{Chen2009} for techniques) and typically the implementation used is from \citet{LH1974}. The orbit weights should sum to $1$ and so $\mathbf{D}$ and 
$\mathbf{K}$ are adjusted to include a sum of weights constraint (a row of $1$s in $\mathbf{D}$ with the corresponding element of $\mathbf{K}$ set to $1$.

The regularised form of expression (\ref{eqn:orbwts}) that we use is
\begin{equation}
	\|\mathbf{Dw} - \mathbf{K}\|^2_2 + \lambda \|\mathbf{w}\|^2_2,
\end{equation}
where $\lambda$ is a user tunable parameter.  This quadratic form of Tikhonov regularisation (\citealt{Tikhonov1963}) maintains the convex form of the expression to be minimised, and acts by suppressing high valued weights (see, for example, \citealt{Vasiliev2013} or \citealt{Valluri2004}).  In this letter, as will be seen in Section \ref{sec:results}, we will use it to increase the number of active orbits in our models and to improve the orbit weight distribution.

As a final point in this subsection, we turn the minimisation into a $\chi^2$ minimisation by dividing the data elements in $\mathbf{K}$ by their errors and similarly the corresponding rows in $\mathbf{D}$.

\subsection{Extensions for modelling Spectral Line Strength Data}\label{sec:enhtheory}
In the previous section, we covered how orbit weights are generated by weighting model observables to match the measured observations.  In this section, we generate the line strength values to be associated with each orbit by asking what model orbit contributions given the orbit weighting will enable the measured values to be matched.  In other words, the roles of the orbit weights and the model's orbit contributions are swapped by comparison with equation (\ref{eqn:orbwts}), with the role of the fractional mass terms being unchanged. Again we will use a non-negative least squares / convex optimisation approach, this time seeking to minimise
\begin{equation}
	\|\mathbf{Cx} - \mathbf{S}\|^2_2 + \lambda_{\rm{LS}} \|\mathbf{x}\|^2_2,
\label{eqn:spec}
\end{equation}
where $\mathbf{C}$ is the `design' matrix giving the individual orbit weightings including the fractional mass terms, $\mathbf{x}$ are the orbit line strength values to be determined,  $\mathbf{S}$ contains the target line strength values for a given spectral line, and $\lambda_{\rm{LS}}$ is the regularisation parameter which may be zero. Only active orbits (orbits with weights non-zero) are included in the design matrix. Note that different expressions (\ref{eqn:spec}) as above exist for each line to be modelled.  For the summation constraint, for the spatial region that we have line strength data, we require that the model sum of orbit line strength values equals the sum of the input line strengths with $\mathbf{C}$ and $\mathbf{S}$ being adjusted accordingly.

Note that, even though we position this letter on modelling line strength data, the above techniques will work on other numerical attributes of orbits provided they are linear in superposition.  Modelling age and metallicity from stellar population analyses could be handled in this way.  Logarithmic data can also be modelled without conversion provided it is acceptable that values will be geometric means not arithmetic means (the arithmetic mean of log data is in fact the log of the geometric mean of the data).

\subsection{Software Implementation}\label{sec:softimpl}
The Python M2M implementation used in \citet{Long2016} has been extended to perform Schwarzschild modelling as well.  Taking this approach means that we can take advantage of existing M2M software concerned with handling initial conditions, gravitational potentials, binning schemes, orbit integration and creation, and parallelisation. We are able to model surface brightness, luminosity density, mean line of sight velocity and mean line of sight velocity squared using Voronoi cells for binning kinematic data. Modelling with Gauss-Hermite coefficients of the line of sight velocity distribution (see \citealt{Rix1997}) has been implemented and used but is not reported on in this letter.  Dithering of orbits is not currently implemented.
Ignoring data preparation, execution of our software takes place in 2 stages.  The first is concerned with orbit creation and collecting orbit data contributing to calculating model observables (the columns of the design matrices $\mathbf{D}$ and $\mathbf{C}$).  The second is concerned with performing the convex optimisations with a variety of different methods and implementations, and different levels of regularisation.  These different methods include non-negative least squares (CVXOPT \footnote{http://cvxopt.org/} and \citet{LH1974} implementations), bounded variable least squares (Python \textit{scipy} implementation), and the sequential coordinate-wise algorithm \citep{Franc2005}. Our preferred optimisation software is CVXOPT which has a parallelised linear algebra package.  Note that the Lawson and Hanson algorithm can be parallelised - see, for example, \citet{Luo2011}.

\section{Application to external galaxies}\label{sec:external}

We use the same ATLAS$^{\rm{3D}}$ data\footnote{http://purl.org/atlas3d} and surface brightness multi-Gaussian expansions (MGEs, see \citealt{Emsellem1994}) for the same galaxies as the chemo-M2M activity in \citet{Long2016}. Considering the constraining observables, the luminosity constraints are as the M2M work but, for the kinematic constraints, we substitute mean velocity squared for the velocity dispersion constraint. We model the same three spectral lines (H$\beta$, Fe5015 and Mg$\,b$) using both symmetrised and unsymmetrised line strength data.  Relevant galaxy properties are shown in Table \ref{tab:galaxyprop}.  If it is needed, the total number of observational constraints contributing to the orbit weights can be determined as 2 X the number of IFU points from Table \ref{tab:galaxyprop} + 512 (for the luminosity constraints).

\begin{table}
	\centering
	\caption{Galaxy Properties}
	\label{tab:galaxyprop}
	\begin{tabular}{ccccc}
		\hline
		Galaxy & Morphology & Inclination & M/L Ratio & IFU data points \\
		\hline
		NGC 1248 & S0 & $42 ^{\circ}$ & $2.50$ &  $297$\\
		NGC 3838 & S0 & $79 ^{\circ}$ & $4.00$ &  $383$\\
		NGC 4452 & S0 & $88 ^{\circ}$ & $5.20$ &  $489$\\
		NGC 4551 & E  & $63 ^{\circ}$ & $4.89$ &  $596$\\
		\hline
	\end{tabular}
		
\medskip
Galaxies and their properties relevant to our Schwarzschild models.  These are same as used in \citet{Long2016} for chemo-M2M modelling.  
\end{table}

We create axisymmetric Schwarzschild models of our galaxies using the MGEs noted above in creating gravitational potentials.  Initial conditions for the orbits are as for the chemo-M2M activity and include the same circularity adjustment for S0 galaxies to create more circular orbits. We use 8000 undithered orbits per galaxy.  Units for modelling are effective radii for length, $10^7$ years for time, and mass in units of the solar mass $M_{\odot}$ with luminosity similarly so.  Line strength data values are in Angstrom.

Both the summation constraints and regularisation involve manually tunable parameters and we use the following parameter values determined by experimentation.  For the summation constraints, the values are $10^3$ for orbit weights, and $1.0$ for orbit line strength contributions.  For regularisation, for orbit weights they are $8 \times 10^{-2}$ for NGC 1248 and $2 \times 10^{-1}$ for the other galaxies, and for line strength $10^{-5}$ for all four galaxies.

\section{Results}\label{sec:results}
It is not practical to show all results for all galaxies using figures.  We therefore focus on a single galaxy NGC 4452 with figures but display results for the other galaxies in tabular form only.  Unless stated otherwise all analyses are conducted using CVXOPT.

\subsection{Initial Models}\label{sec:initmodels}
\begin{figure*}
\centering
\begin{tabular}{lcr}
\includegraphics[width=45mm]{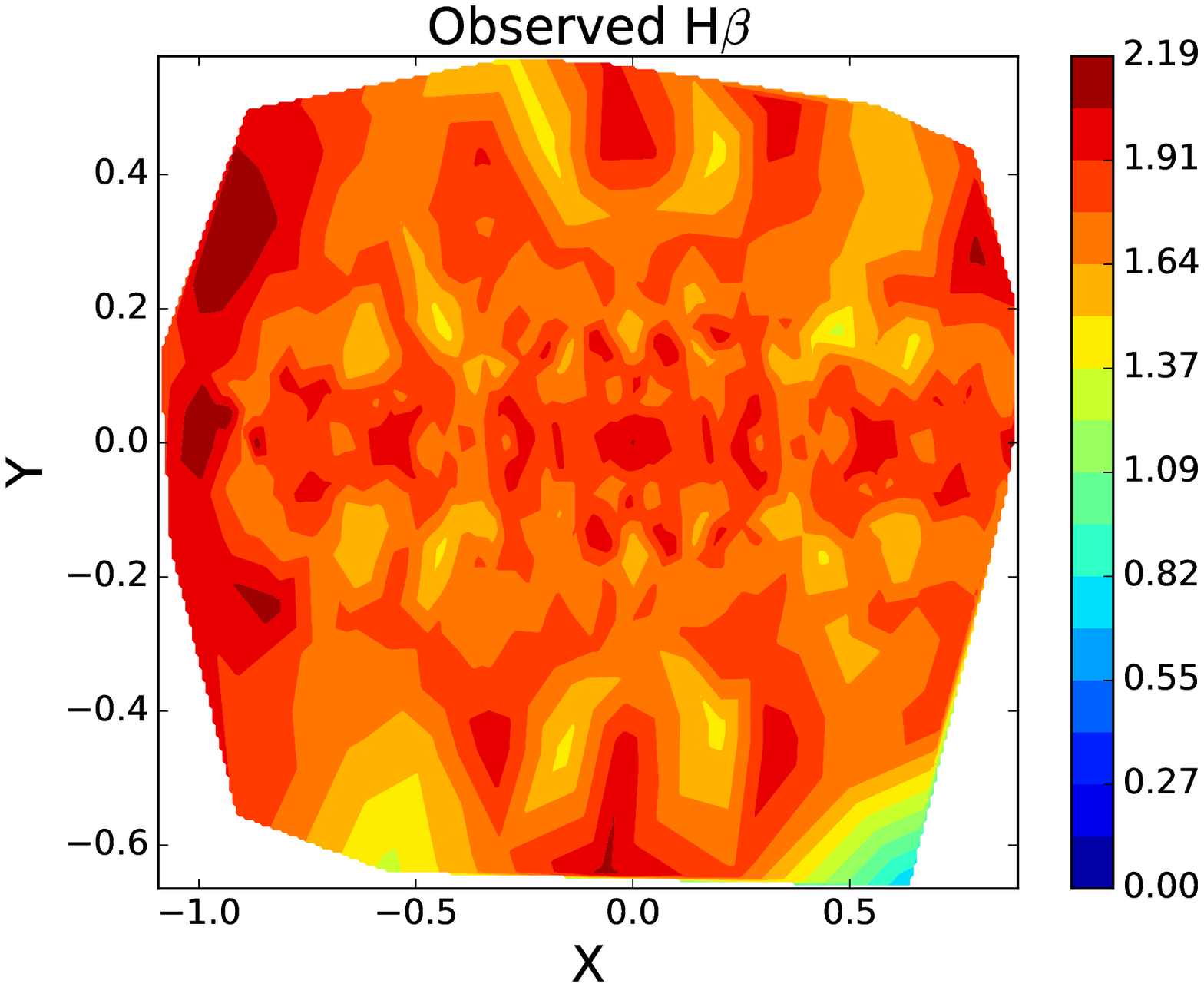} & \includegraphics[width=45mm]{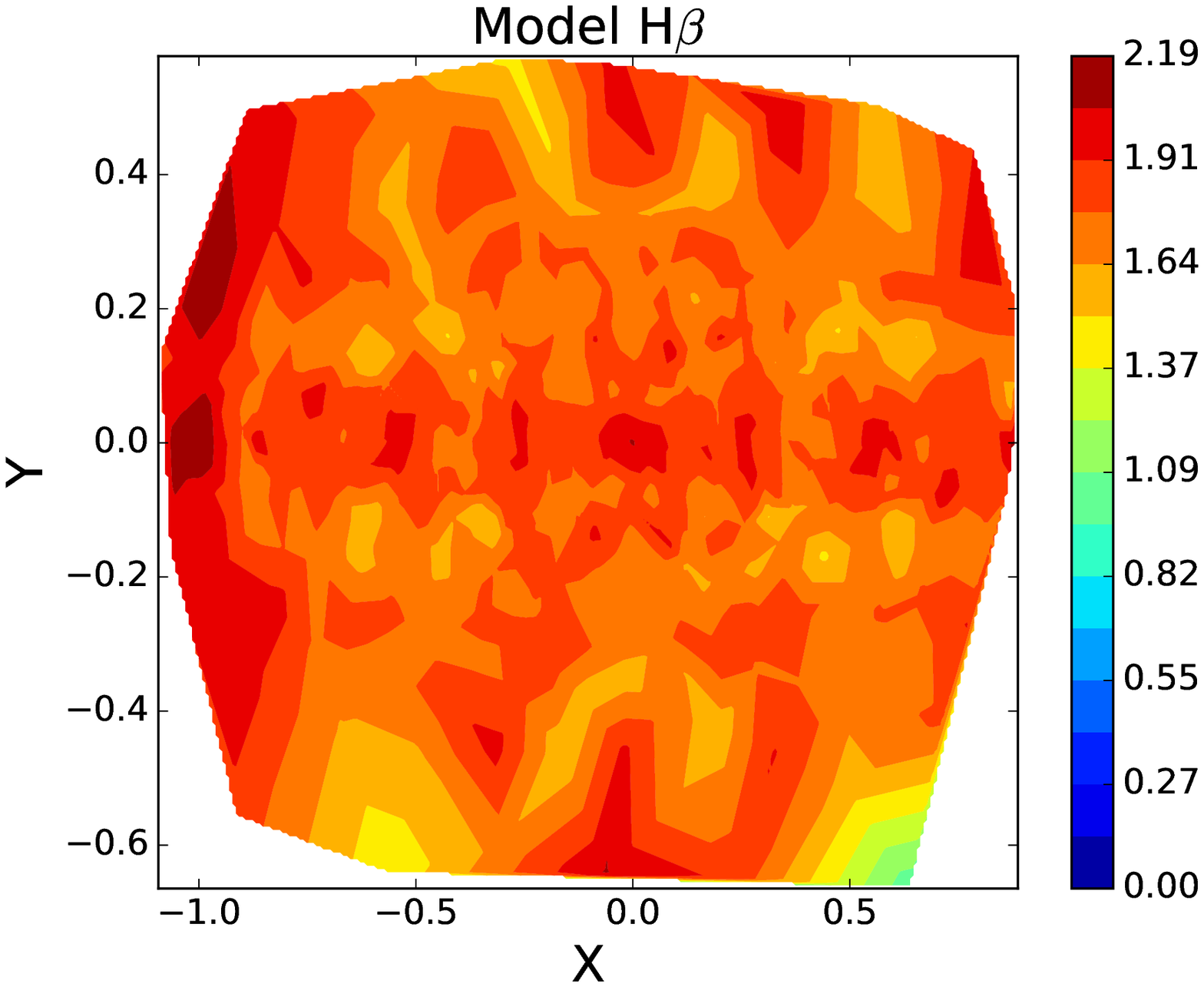}  & \includegraphics[width=45mm]{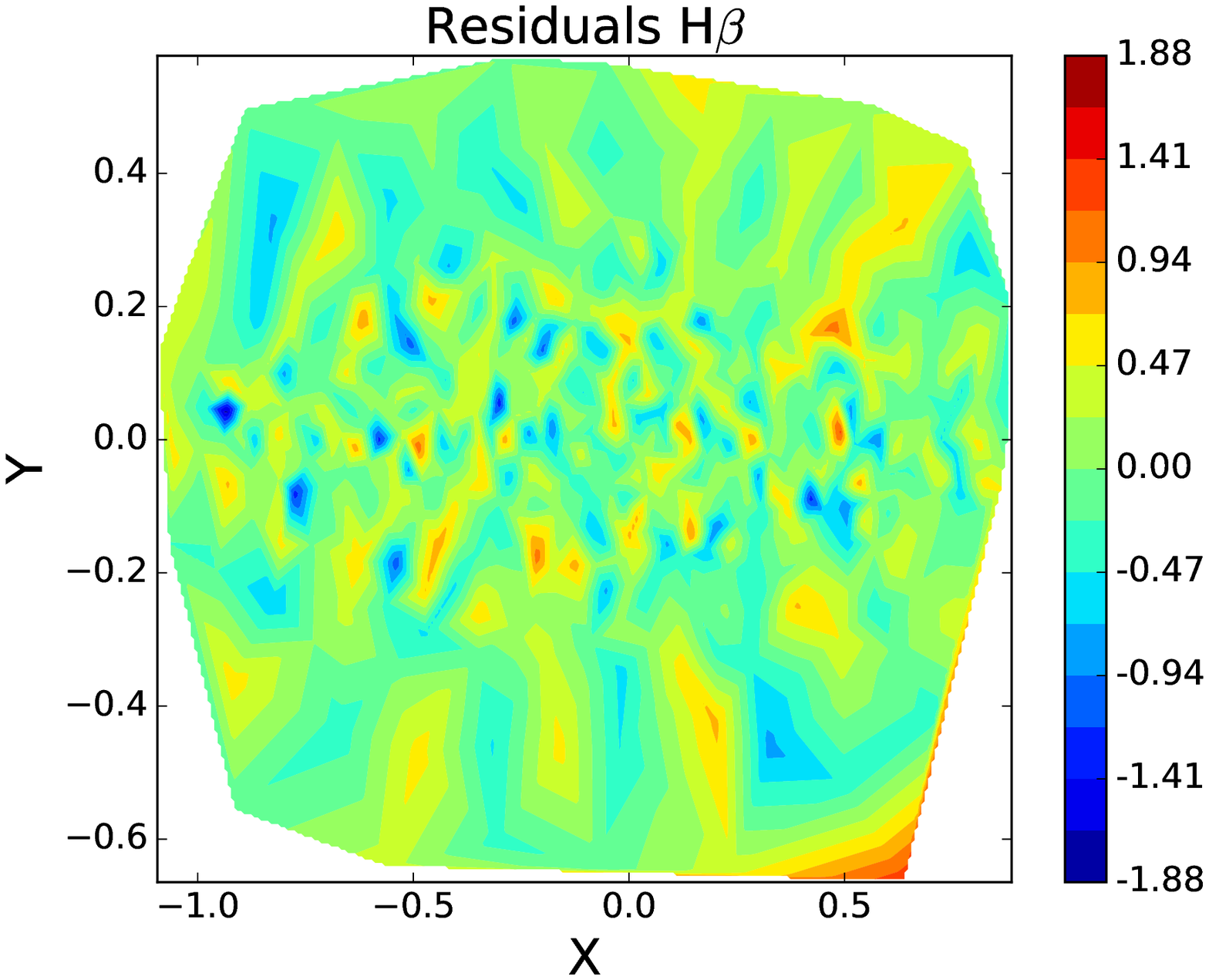}\\
\includegraphics[width=45mm]{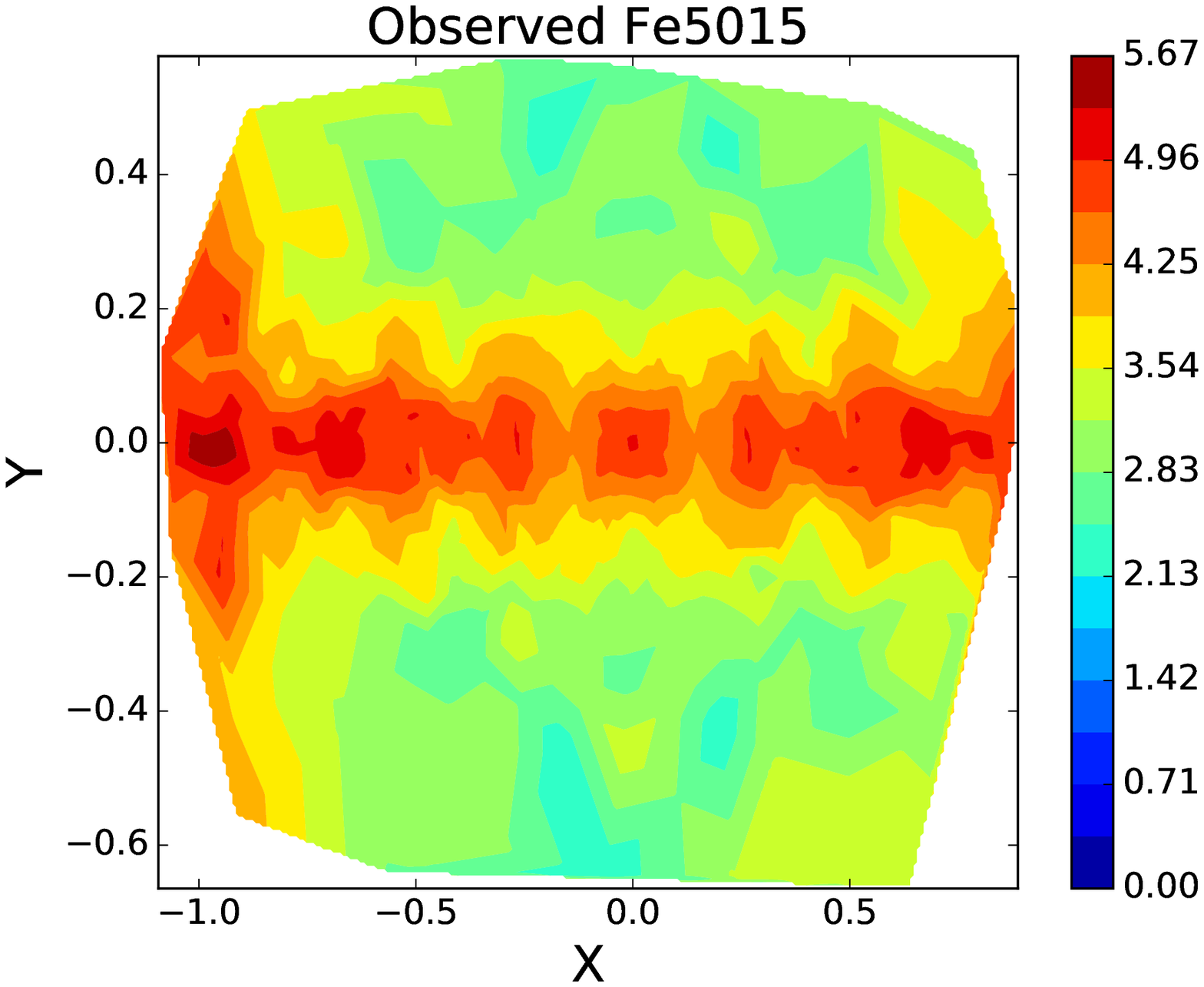} & \includegraphics[width=45mm]{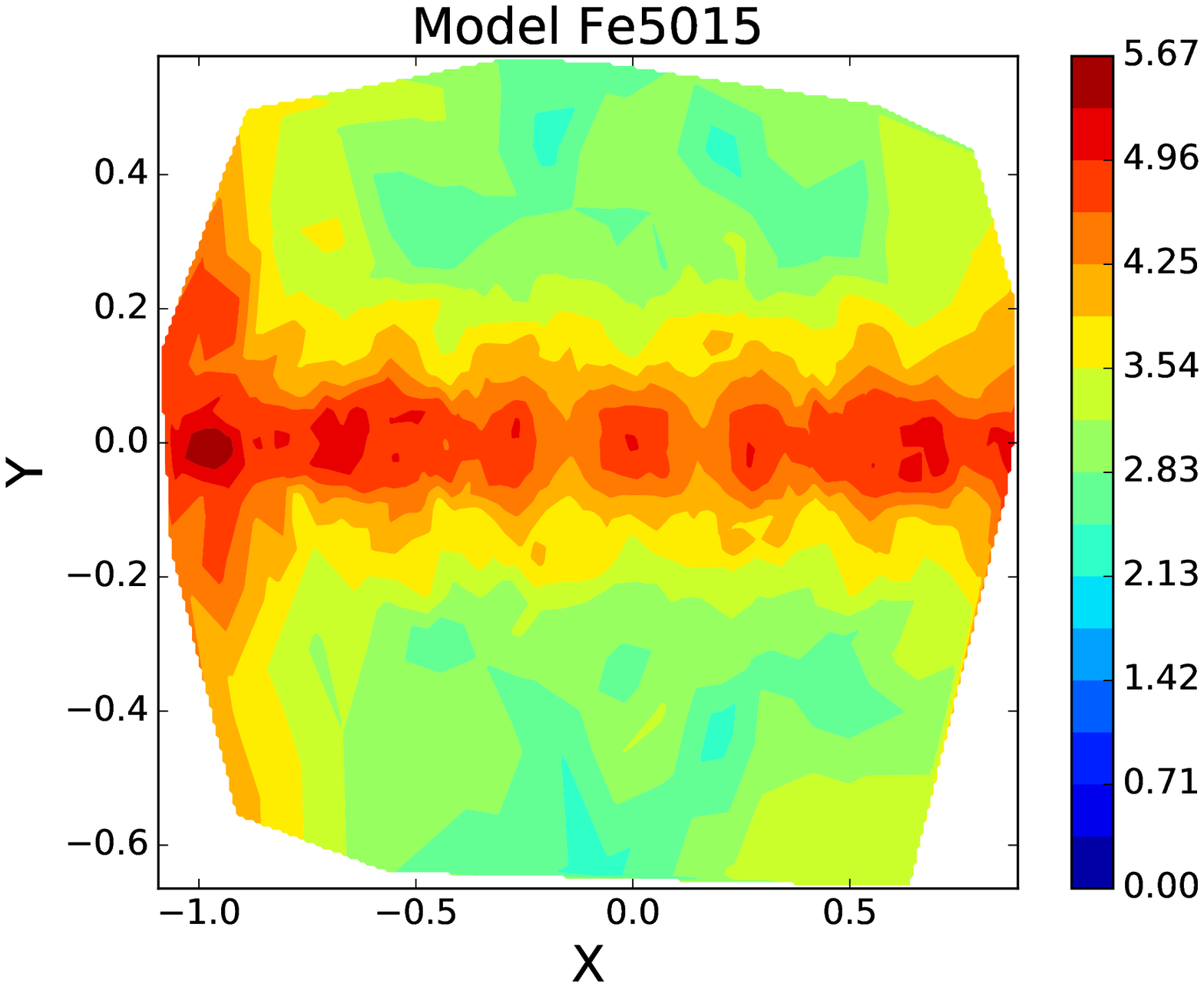}  & \includegraphics[width=45mm]{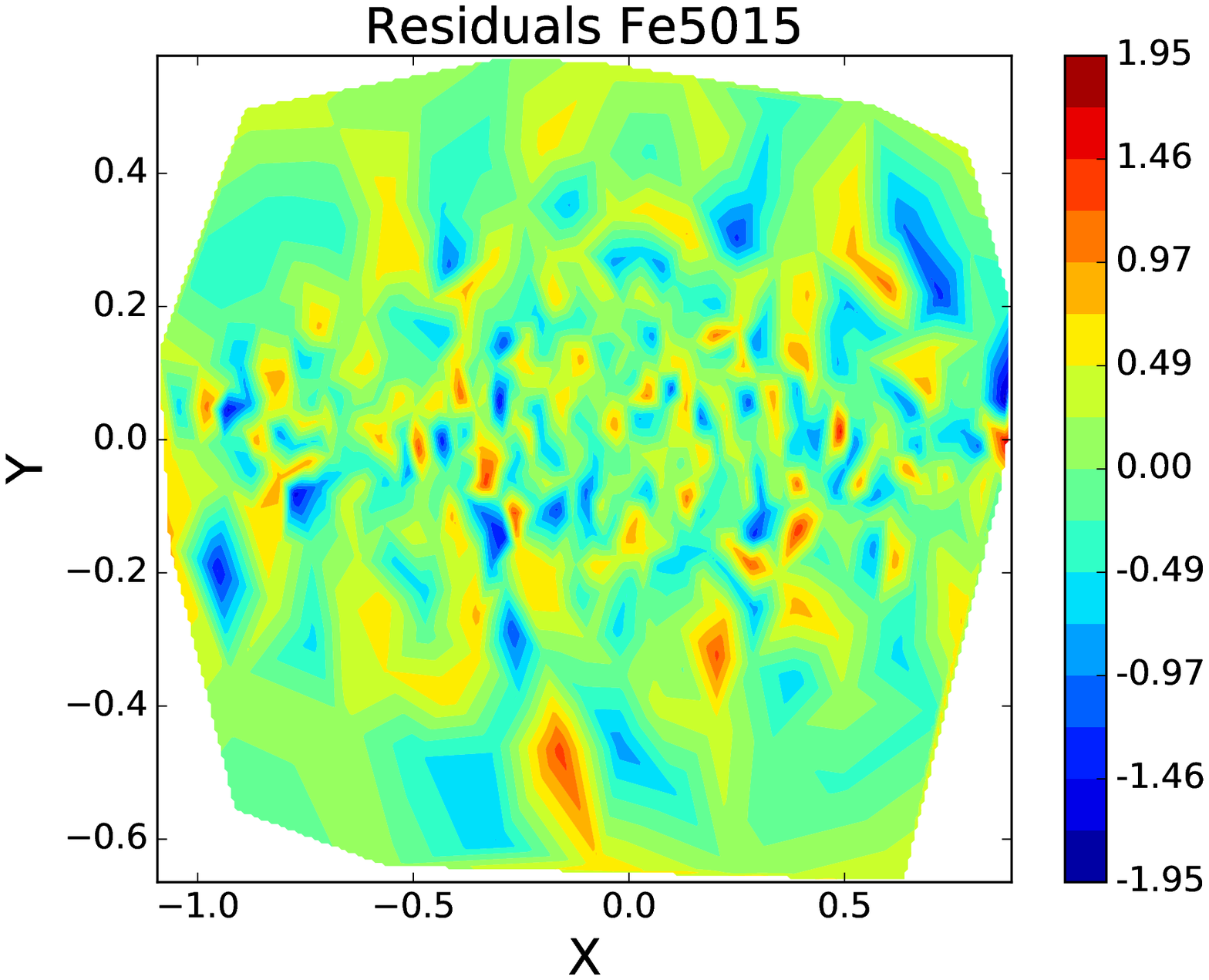}\\
\includegraphics[width=45mm]{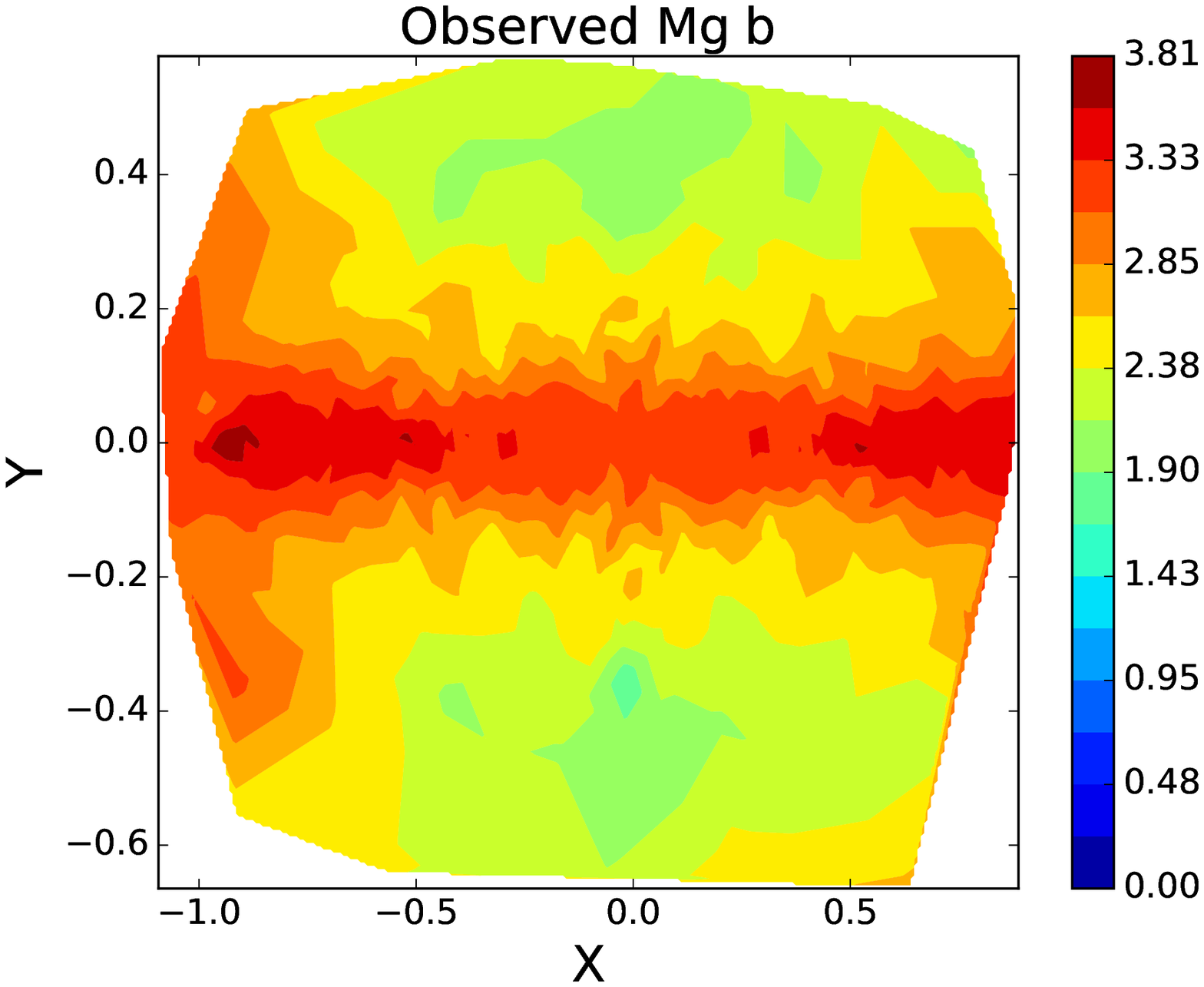} & \includegraphics[width=45mm]{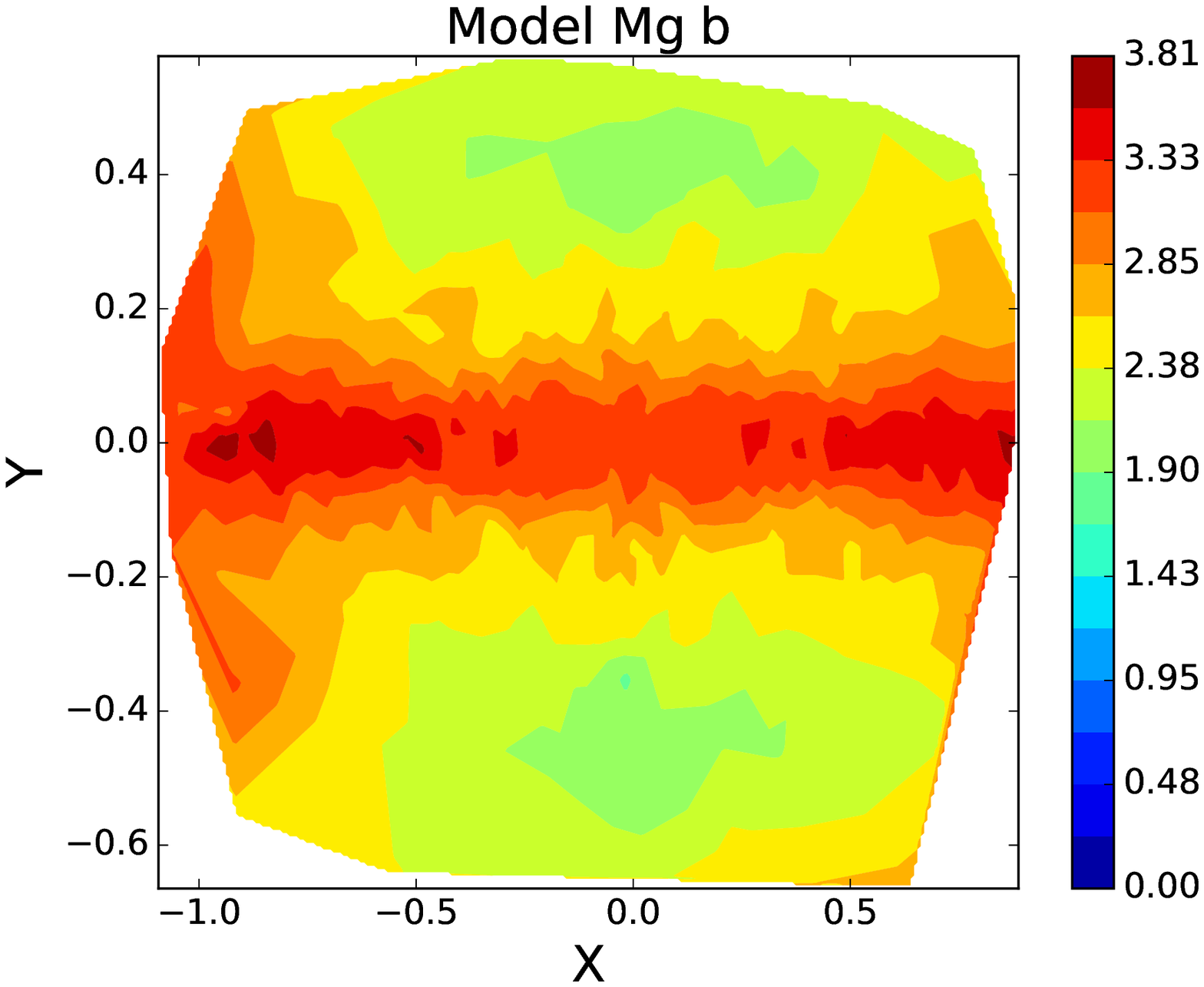}  & \includegraphics[width=45mm]{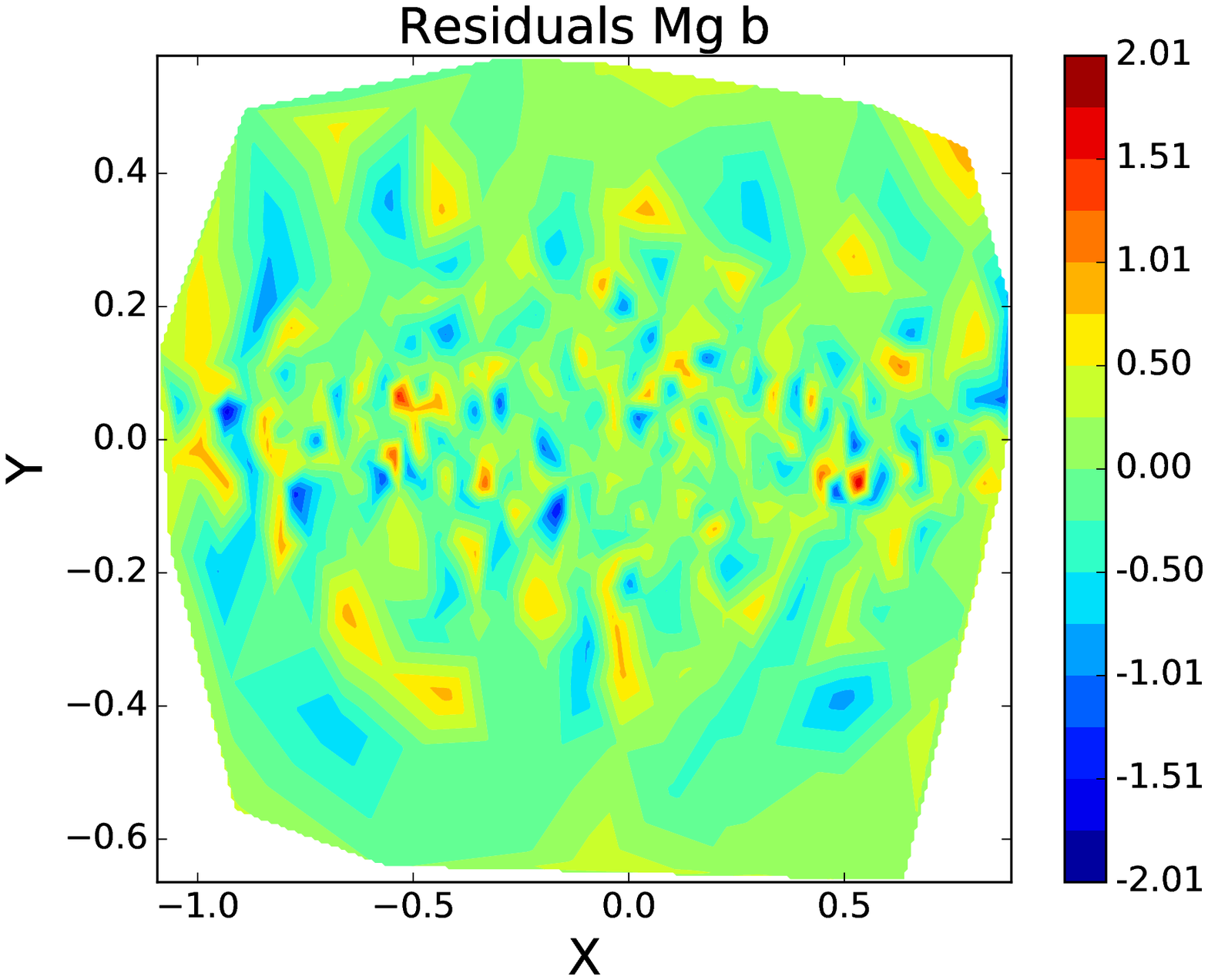}\\
\includegraphics[width=45mm]{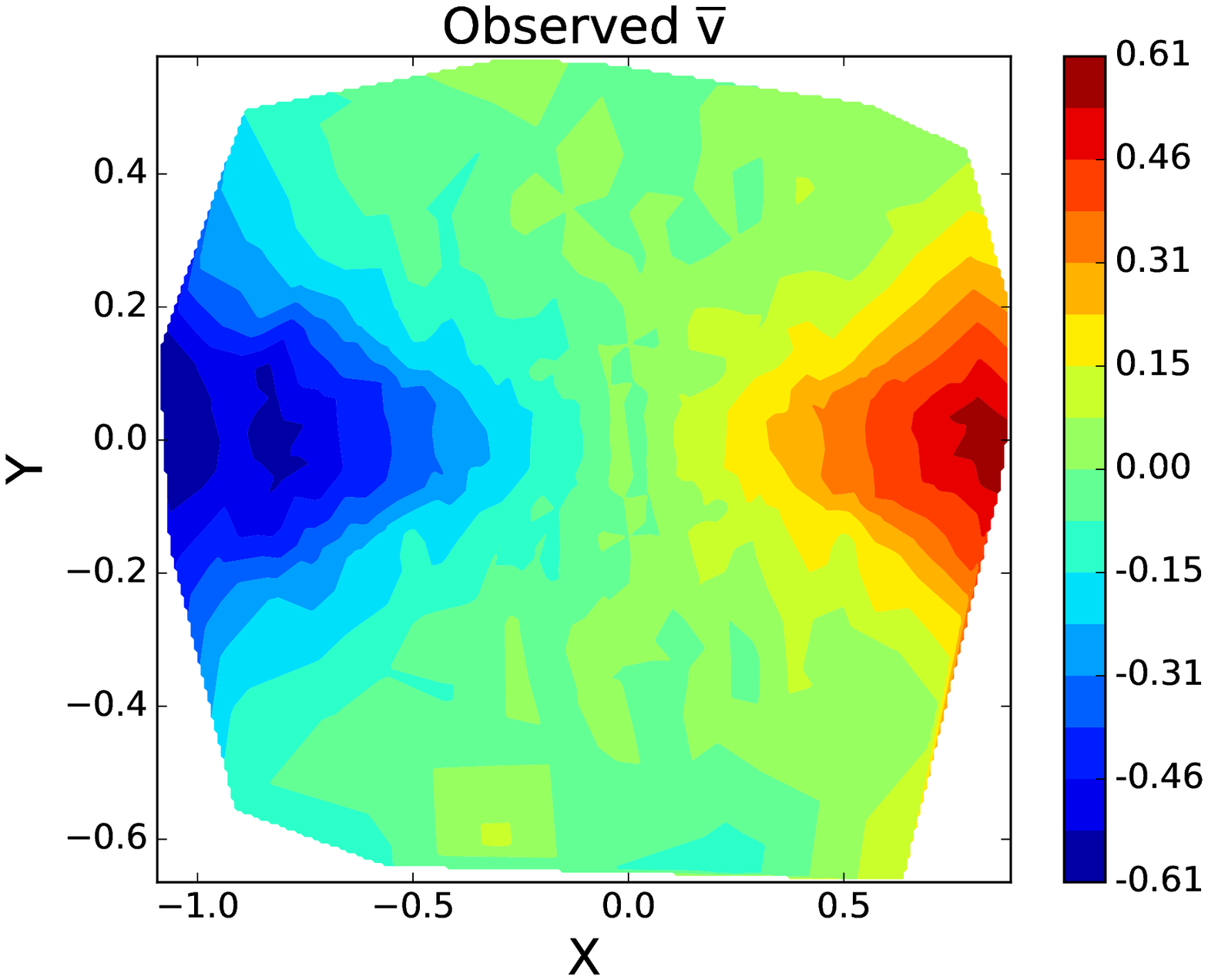} & \includegraphics[width=45mm]{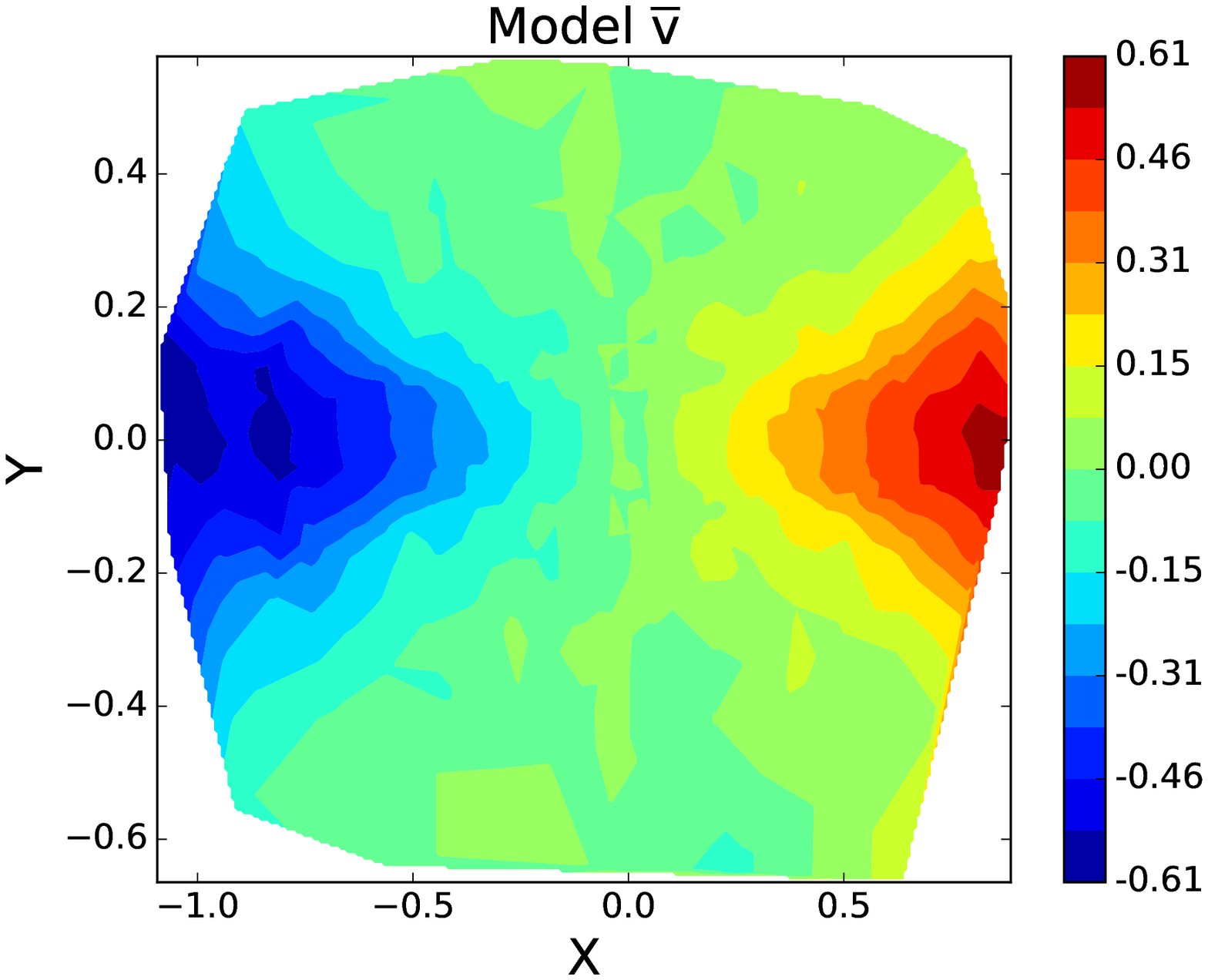}  & \includegraphics[width=45mm]{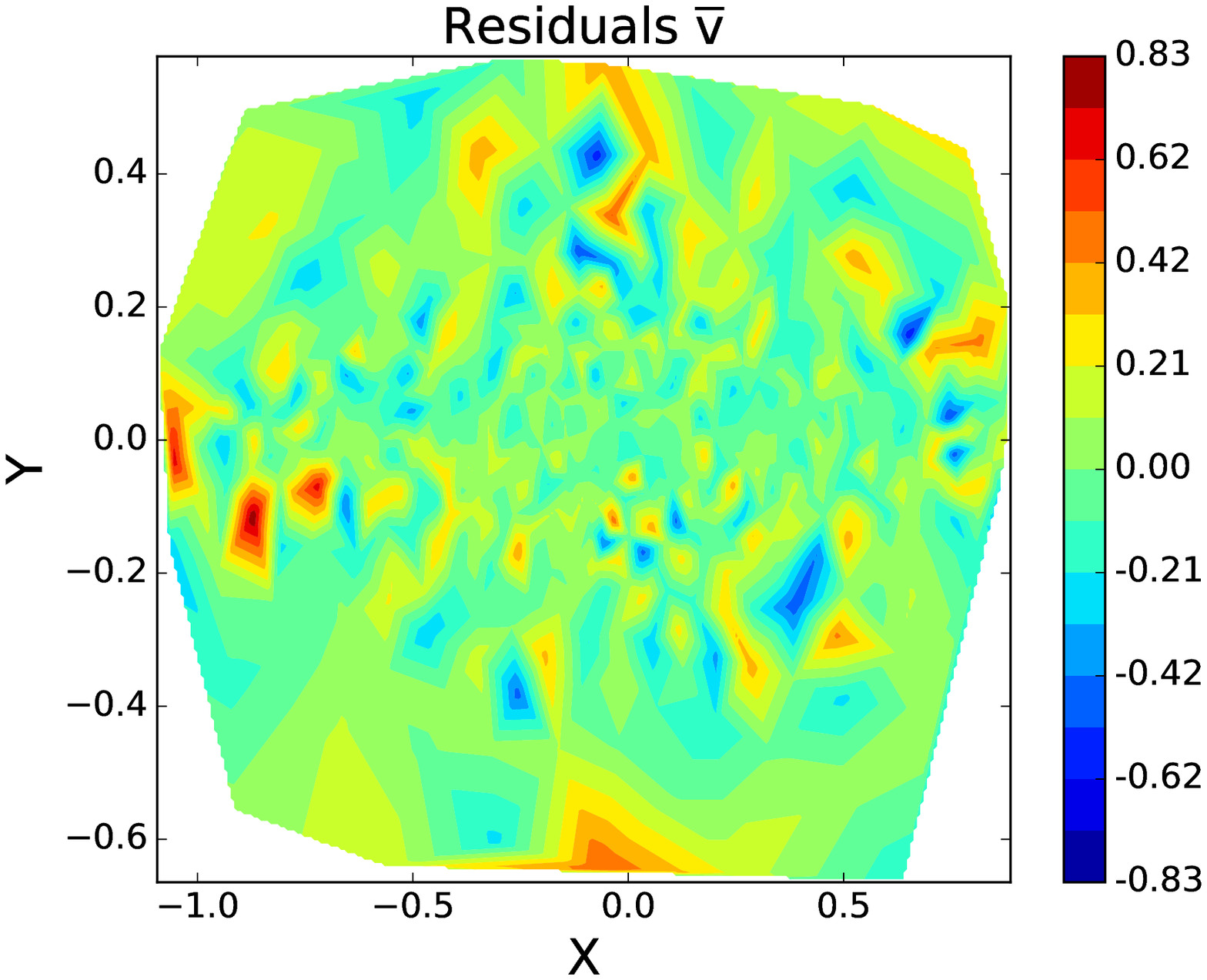}\\
\includegraphics[width=45mm]{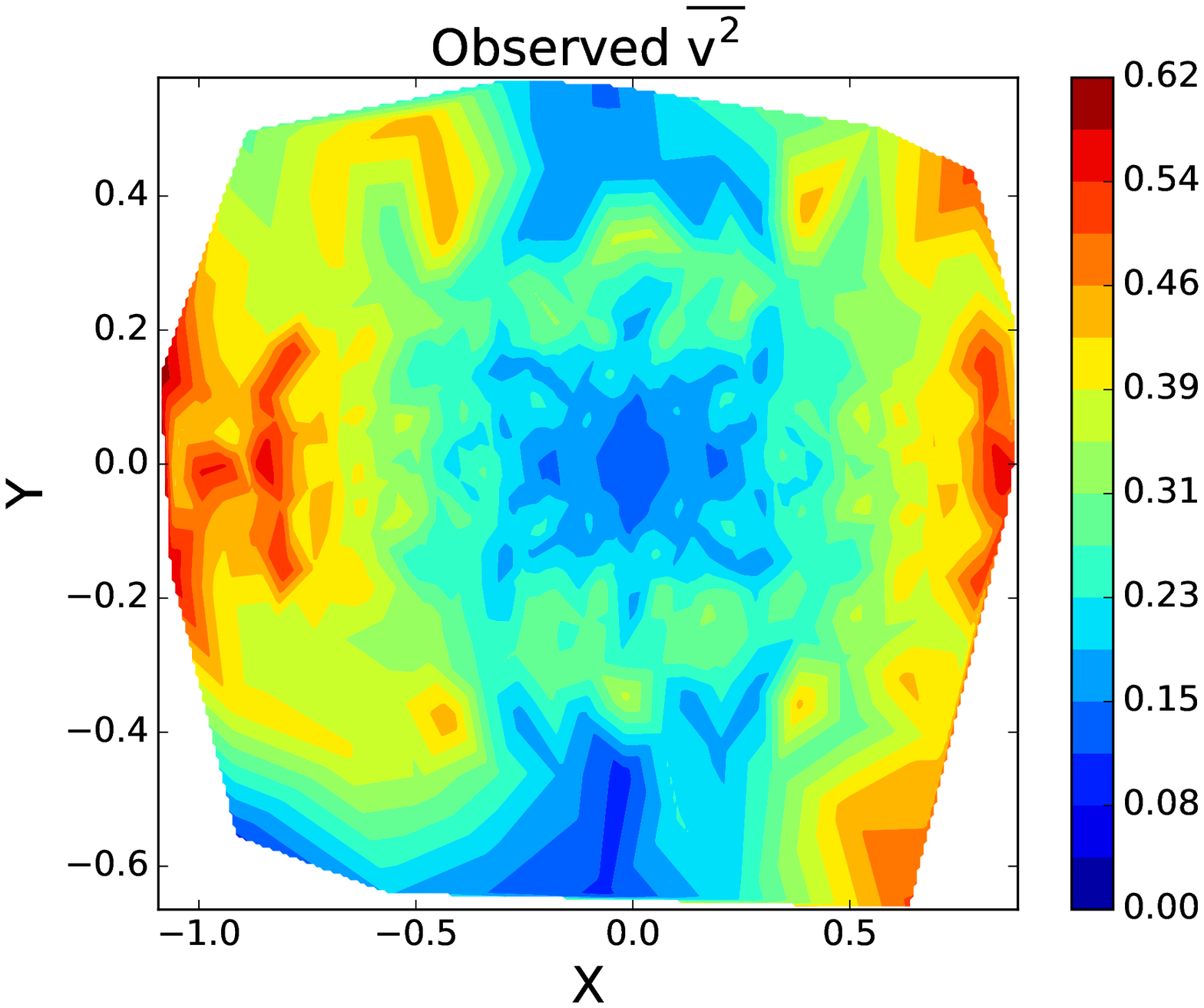} & \includegraphics[width=45mm]{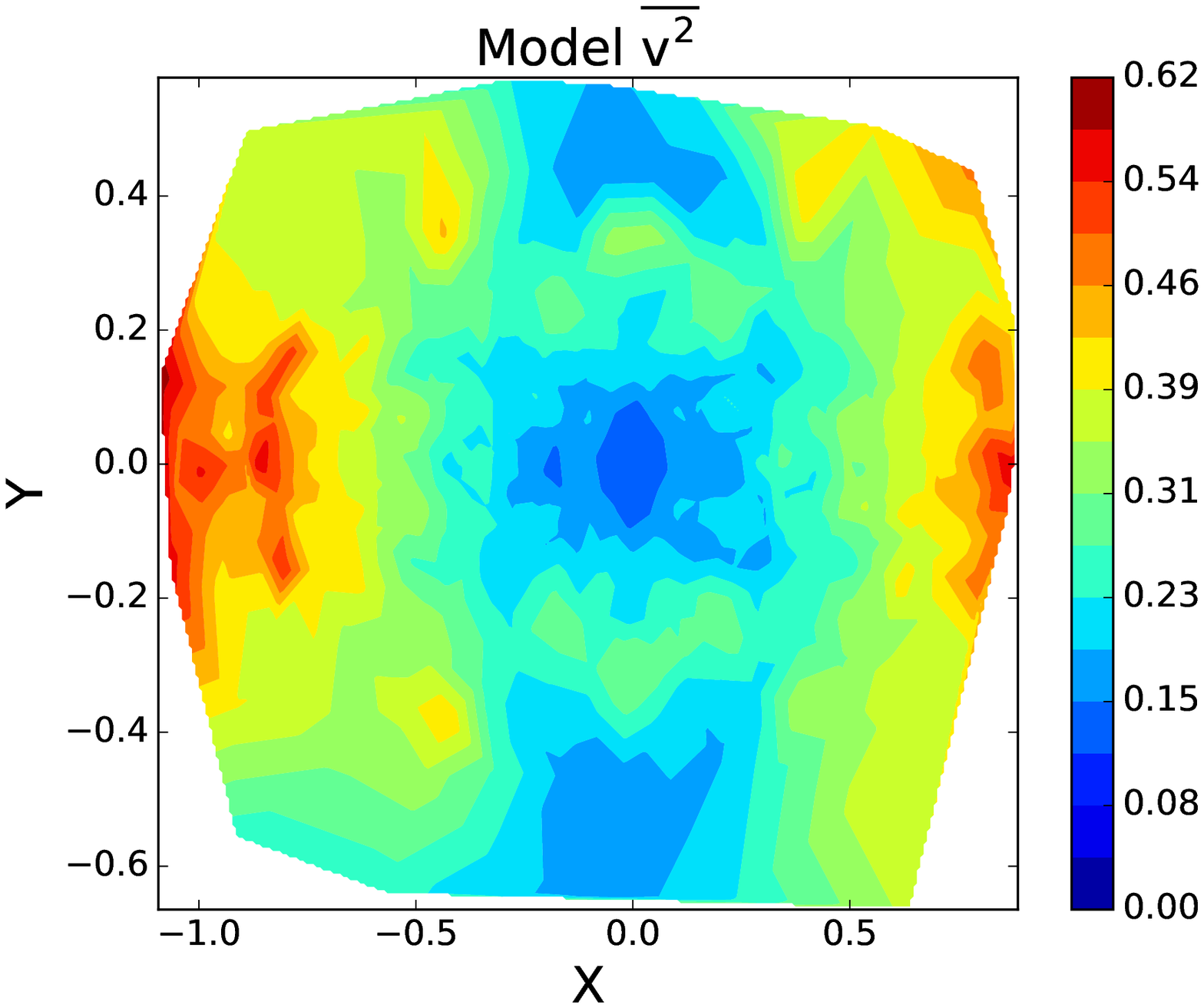}  & \includegraphics[width=45mm]{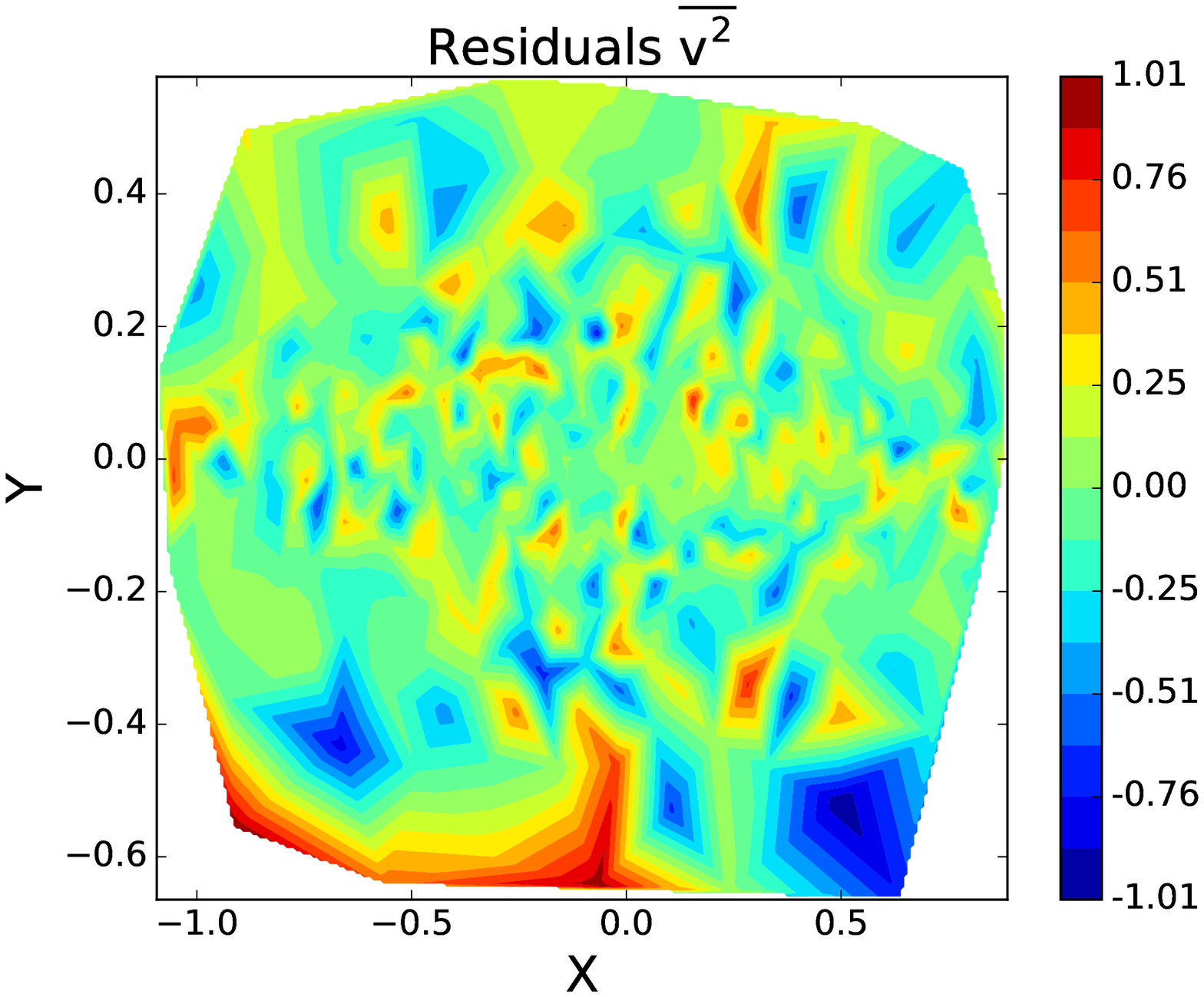}\\
\end{tabular}

\medskip
\caption{NGC 4452 line strength and kinematic plots from chemo-Schwarzschild modelling. Units are as per Section \ref{sec:external}.  Coordinates X and Y give on-sky positions in effective radii.} 
\label{fig:NGC4452}
\end{figure*}

Our initial models for the four galaxies do not use regularisation in the weight determinations. We show the mean $\chi ^2$ values we achieve in Table \ref{tab:cvxoptchi2} top rows, and plots specific to NGC 4452 in Figure \ref{fig:NGC4452}.  It should be clear from the table and figure that we are able to model line strength data as well as luminosity and kinematic data.  All the data in these inital models have been symmetrised.

\begin{table}
	\centering
	\caption{CVXOPT - Mean $\chi ^2$ Values}
	\label{tab:cvxoptchi2}
	\begin{tabular}{lccccccc}
		\hline
		Galaxy & SB & LD & $\bar{v}$ & $\bar{v^2}$ & H$\beta$ & Mg$\,b$ & Fe5015 \\
		\hline
		Without regularisation \\
		NGC 1248 & 0.062 & 0.173 & 0.030 & 0.080 & 0.398 & 0.498 & 0.497 \\
		NGC 3838 & 0.087 & 0.365 & 0.023 & 0.037 & 0.399 & 0.174 & 0.337 \\
		NGC 4452 & 0.071 & 0.024 & 0.040 & 0.087 & 0.219 & 0.267 & 0.362 \\
		NGC 4551 & 0.040 & 0.075 & 0.032 & 0.065 & 0.436 & 0.267 & 0.380 \\
		\hline
		With regularisation \\
		NGC 1248 & 0.063 & 0.296 & 0.059 & 0.318 & 0.193 & 0.185 & 0.807 \\
		NGC 3838 & 0.178 & 1.007 & 0.144 & 0.369 & 0.307 & 0.128 & 0.340 \\
		NGC 4452 & 0.436 & 0.117 & 0.189 & 0.286 & 0.207 & 0.194 & 0.428 \\
		NGC 4551 & 0.147 & 0.313 & 0.068 & 0.356 & 0.301 & 0.215 & 0.569 \\
		\hline
	\end{tabular}
		
\medskip
Mean $\chi ^2$ values calculated using CVXOPT with and without regularisation. SB indicates surface brightness and LD, luminosity density.
\end{table}

\begin{figure*}
\centering
\begin{tabular}{cc}
\includegraphics[width=50mm]{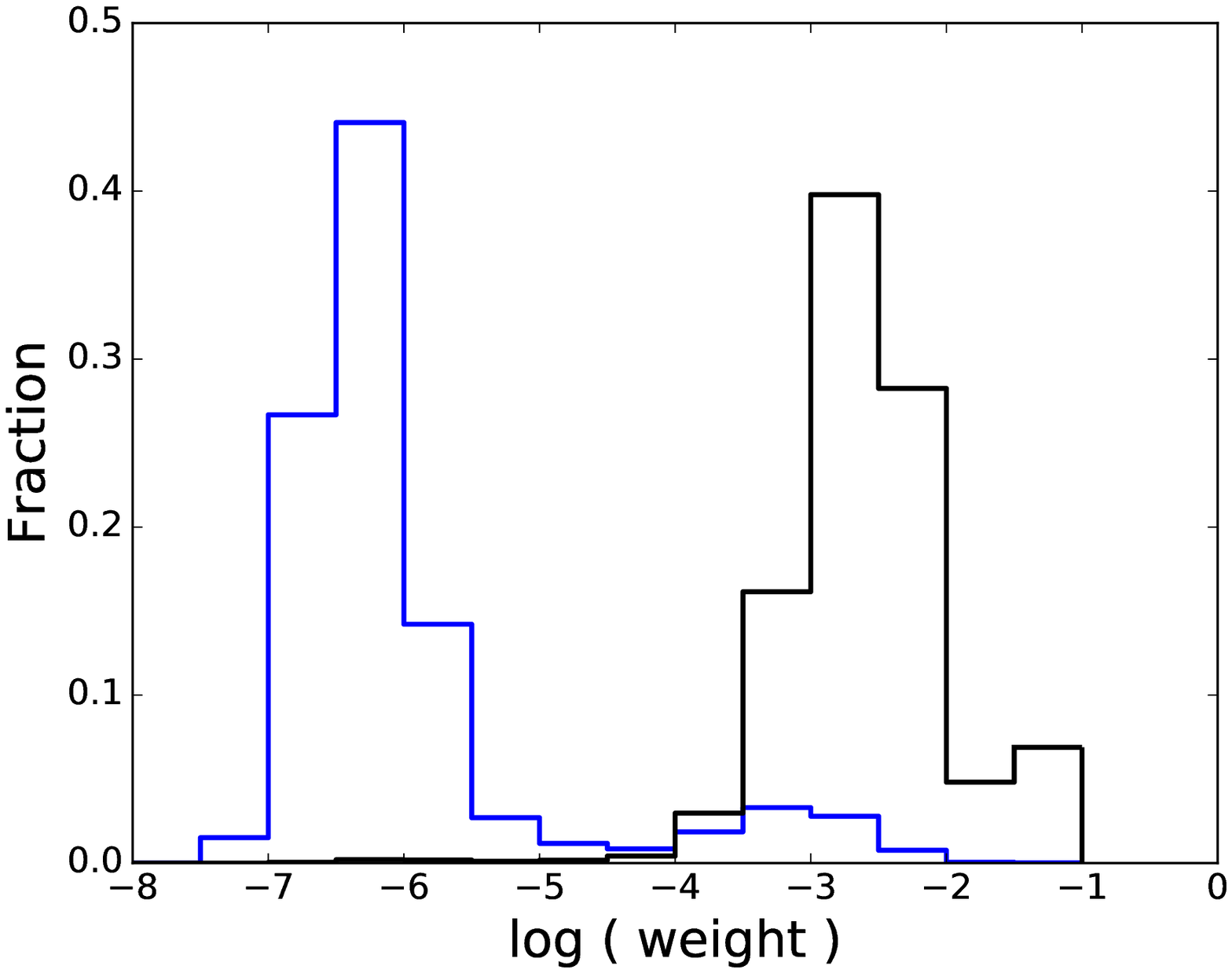} &  \includegraphics[width=50mm]{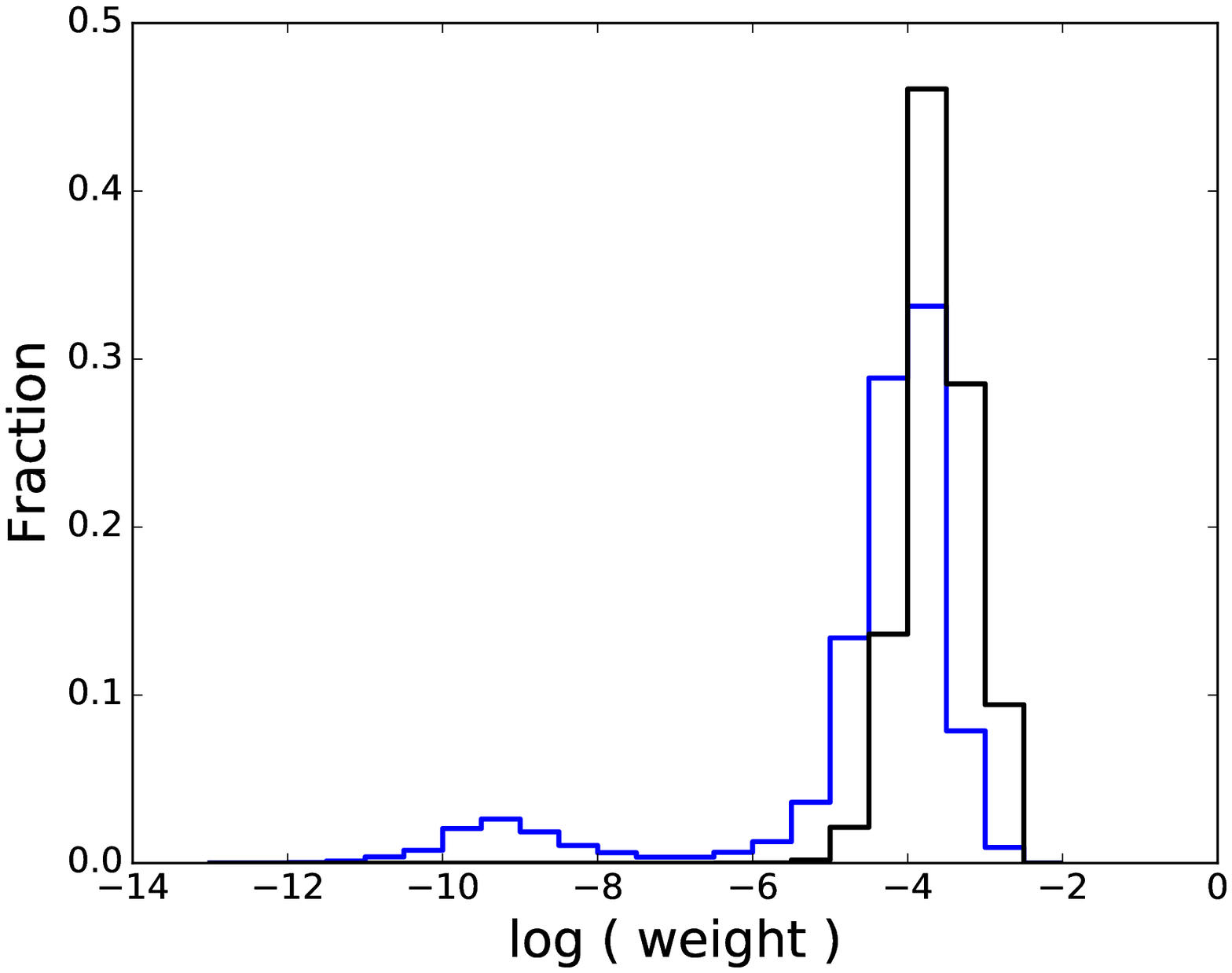}\\
\end{tabular}

\medskip
\caption{NGC 4452 orbit weight distribution with and without regularisation. The left panel (no regularisation) shows that larger orbit weights (99\% of total weight - black line) are associated with a small number of orbits (10\% - blue line) and that many orbits (90\% - blue line) have low orbit weights.  The black line indicates the fractional number of orbits per logarithmic weight bin and the blue line, the fractional weight contained in each bin.  Using regularisation (right panel), the imbalance is reduced with more than 99\% of the total orbit weight distributed across 84\% of the orbits.} 
\label{fig:NGC4452orbwts}
\end{figure*}

We examine quite how the enhanced modelling scheme is behaving.  The sum of weights constraints are met without no problems encountered. For example, the model sum of orbit weights is 1.00.  As can be seen from Figure \ref{fig:NGC4452orbwts} (left panel) for NGC 4452, the orbit weight distribution shows that larger orbit weights (99\% of total weight) are associated with a small fraction of orbits (10\% of orbits) and that many orbits (90\%) have low orbit weights.  This issue is not specific to NGC 4452 and applies to the other three galaxies as well.

If we now include regularisation in our models, as can be seen from Figure \ref{fig:NGC4452orbwts} (right panel), the imbalance in the left panel is significantly reduced and the number of active orbits increases considerably (from 10\% to 84\%).  The mean $\chi ^2$ values we achieve are shown in Table \ref{tab:cvxoptchi2} bottom rows.  As before, the sum of weights constraints are met.

\subsection{Minimisation methods}\label{sec:lhmodels}
In Section \ref{sec:initmodels}, we used the CVXOPT software to analyse our models.  This software uses what is known as an `interior point' method to determine the orbit weights and line strength contributions.  The \citet{LH1974} NNLS method uses an `active set' method and deliberately zeroises weights and contributions for orbits it decides it derives no benefit from using.  For convex function minimisation, if a solution exists, it is unique (this comes directly from the definition of a convex function).   This means that, to within some numerical accuracy, the CVXOPT software and the \citet{LH1974} NNLS method should be returning the same values for the weights and contributions.  We re-analyse our models using \citet{LH1974} and compare the results obtained with the CVXOPT results.  The \citet{LH1974} mean $\chi^2$ values are in Table \ref{tab:LHchi2}.  To three decimal places, the luminosity and kinematic $\chi^2$ values are the same as those in Table \ref{tab:cvxoptchi2} while the line strength values differ due to different numbers of active orbits.  
\begin{table}
	\centering
	\caption{\citealt{LH1974} - Mean $\chi ^2$ Values}
	\label{tab:LHchi2}
	\begin{tabular}{lccccccc}
		\hline
		Galaxy & SB & LD & $\bar{v}$ & $\bar{v^2}$ & H$\beta$ & Mg$\,b$ & Fe5015 \\
		\hline
		Without regularisation \\
		NGC 1248 & 0.062 & 0.173 & 0.030 & 0.080 & 0.370 & 0.463 & 0.474 \\
		NGC 3838 & 0.087 & 0.365 & 0.023 & 0.037 & 0.362 & 0.131 & 0.300 \\
		NGC 4452 & 0.071 & 0.024 & 0.040 & 0.087 & 0.220 & 0.220 & 0.309 \\
		NGC 4551 & 0.040 & 0.075 & 0.032 & 0.065 & 0.429 & 0.264 & 0.354 \\
		\hline
		With regularisation \\
		NGC 1248 & 0.063 & 0.296 & 0.059 & 0.318 & 0.125 & 0.115 & 0.654 \\
		NGC 3838 & 0.178 & 1.007 & 0.144 & 0.369 & 0.235 & 0.075 & 0.255 \\
		NGC 4452 & 0.436 & 0.117 & 0.189 & 0.286 & 0.191 & 0.170 & 0.381 \\
		NGC 4551 & 0.147 & 0.313 & 0.068 & 0.356 & 0.226 & 0.135 & 0.454 \\
		\hline	
	\end{tabular}
		
\medskip
Mean $\chi ^2$ values calculated using \citet{LH1974} with and without regularisation.  The luminosity and kinematic values agree with those obtained using CVXOPT - see Table \ref{tab:cvxoptchi2}.  The line strength values do not due to differing numbers of active orbits.
\end{table}

\begin{figure*}
\centering
\begin{tabular}{cccc}
	Orbit weights & H$\beta$ & Mg$\,b$ & Fe5015\\
\includegraphics[width=35mm]{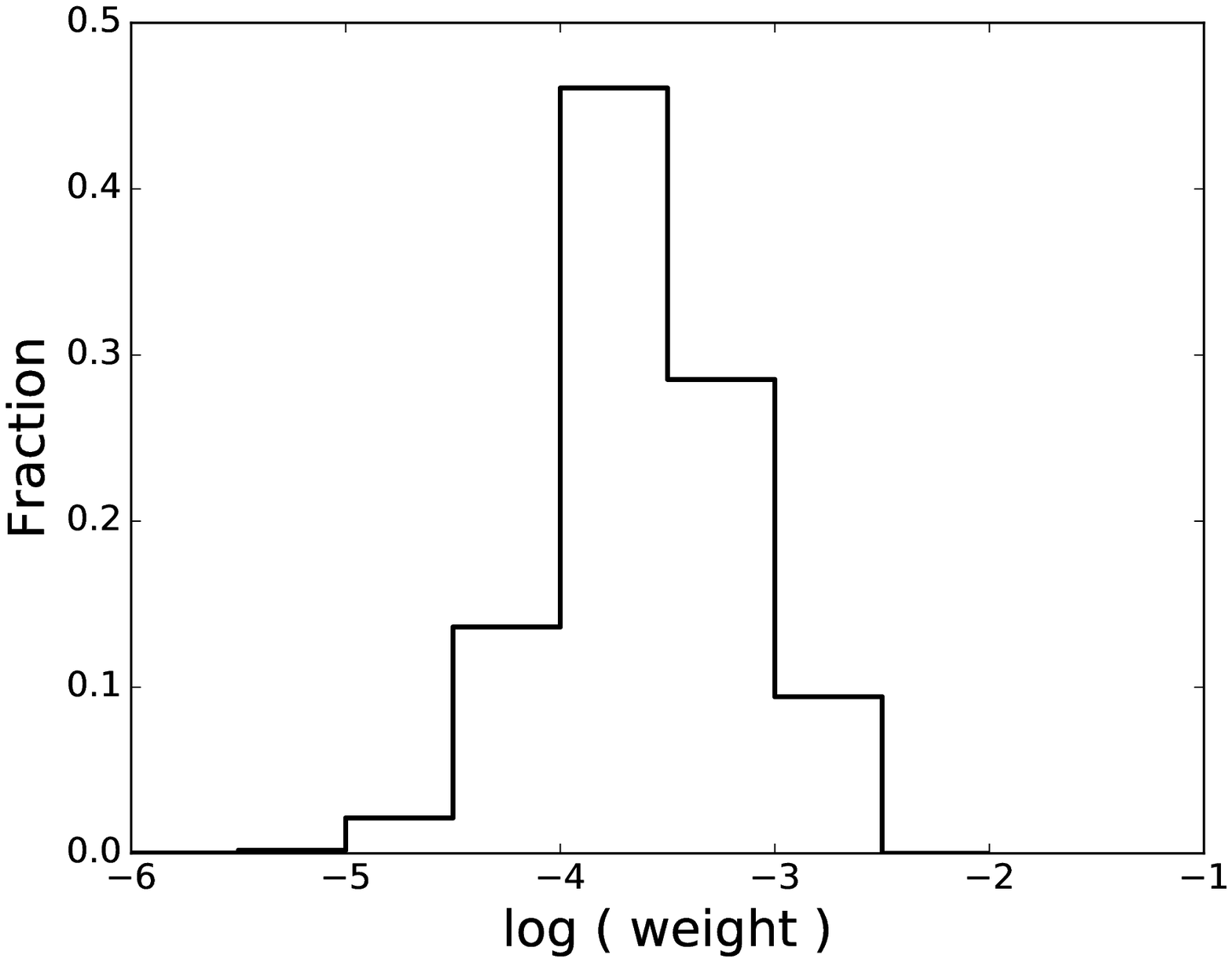} & \includegraphics[width=35mm]{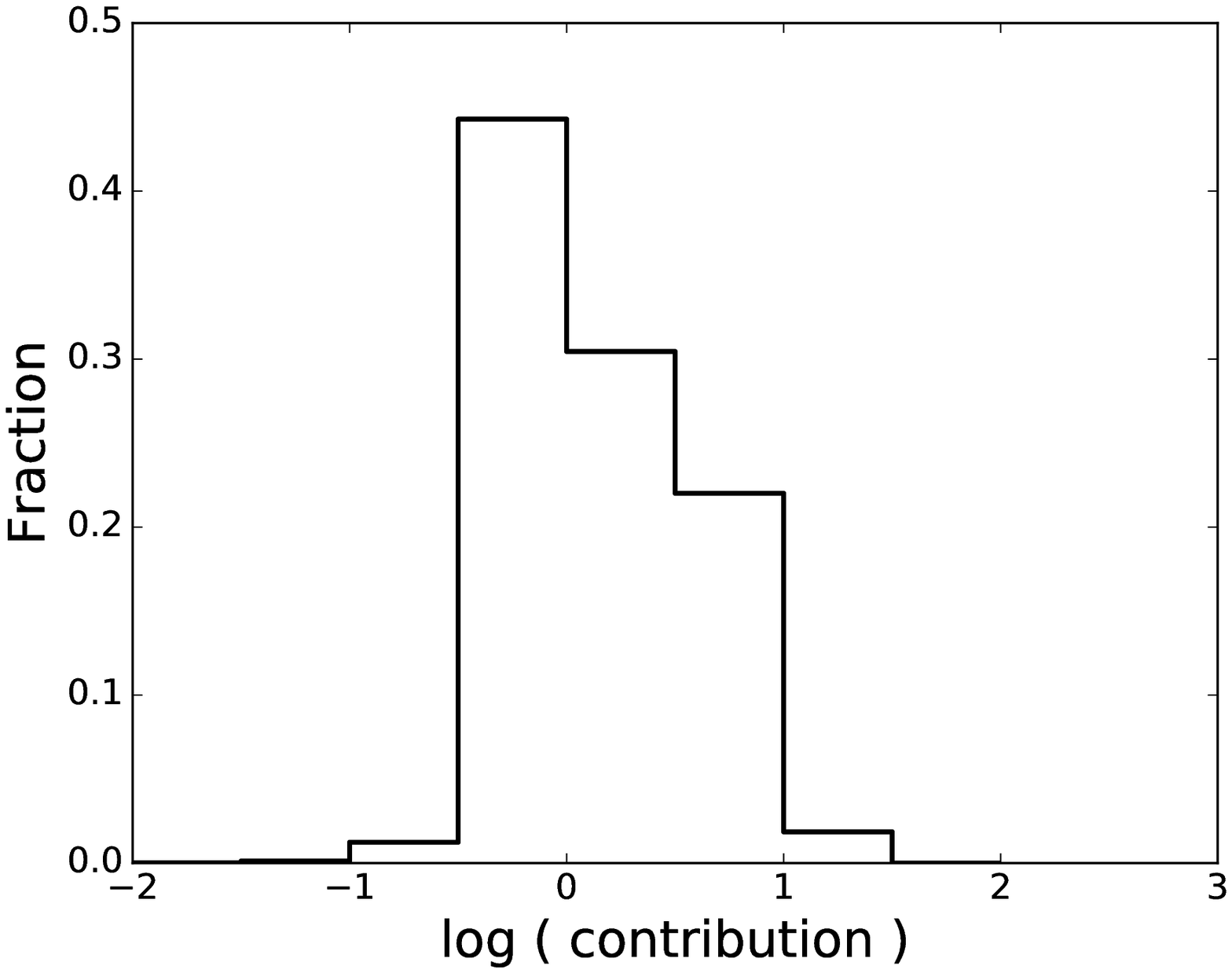} & \includegraphics[width=35mm]{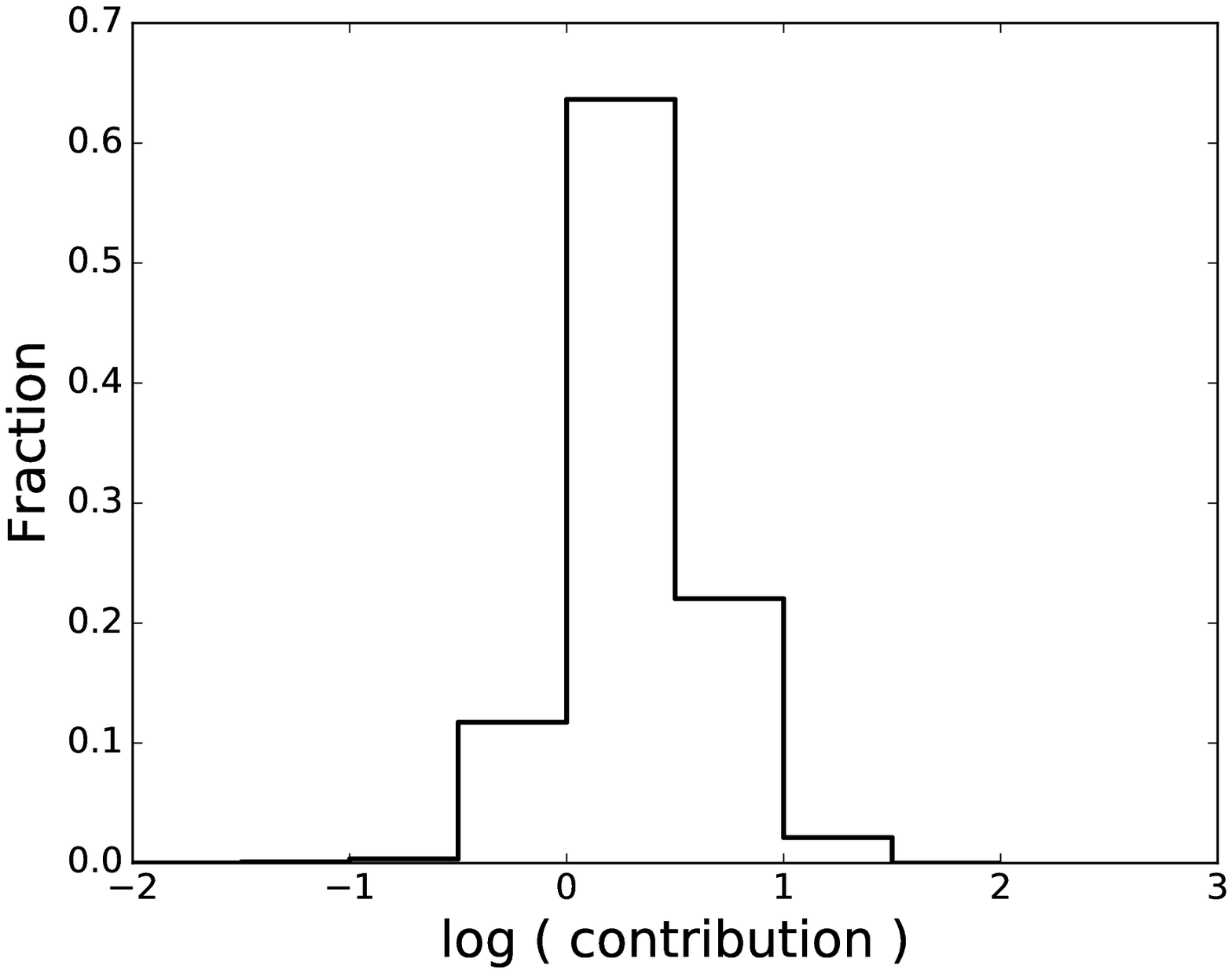} & \includegraphics[width=35mm]{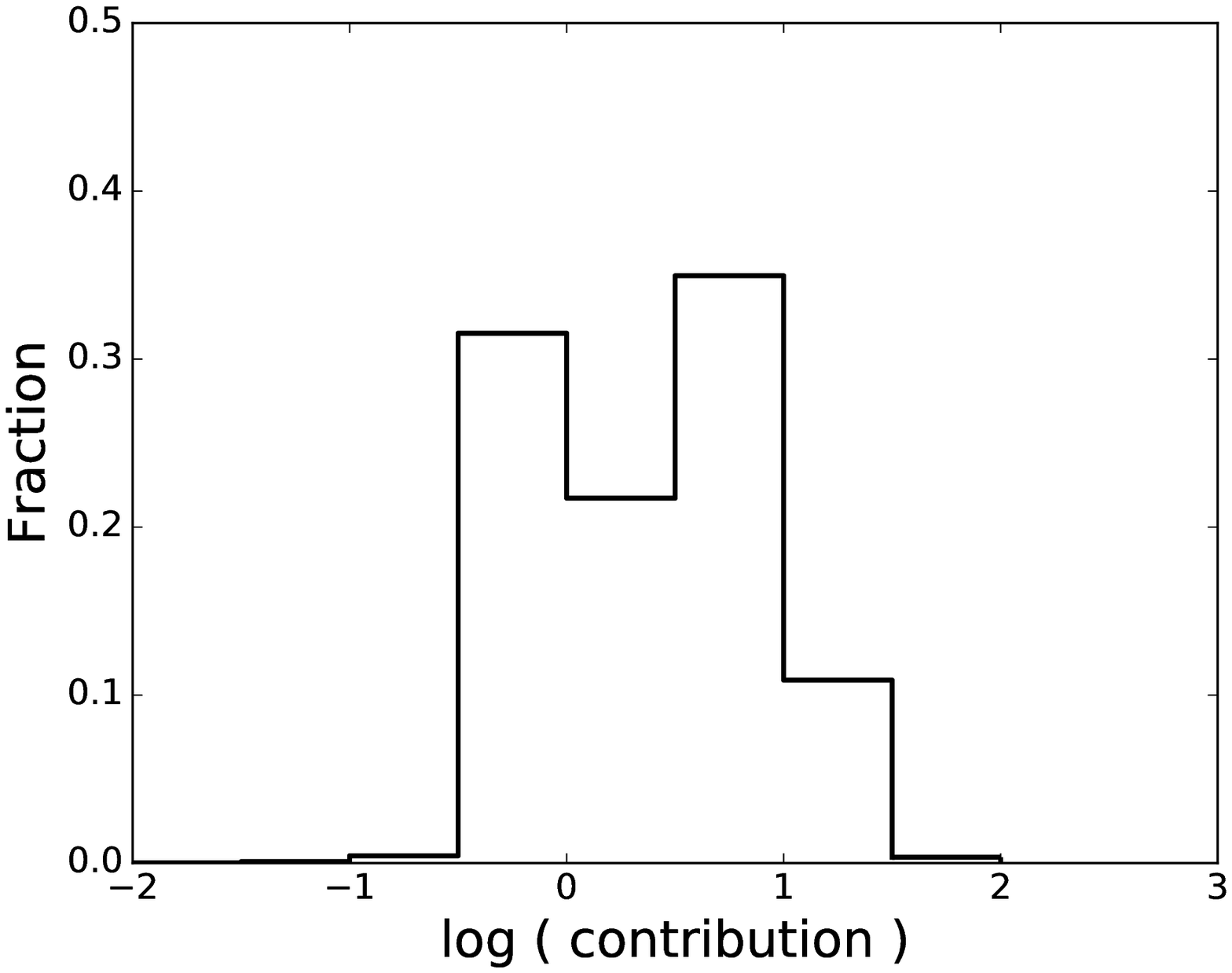} \\
\includegraphics[width=35mm]{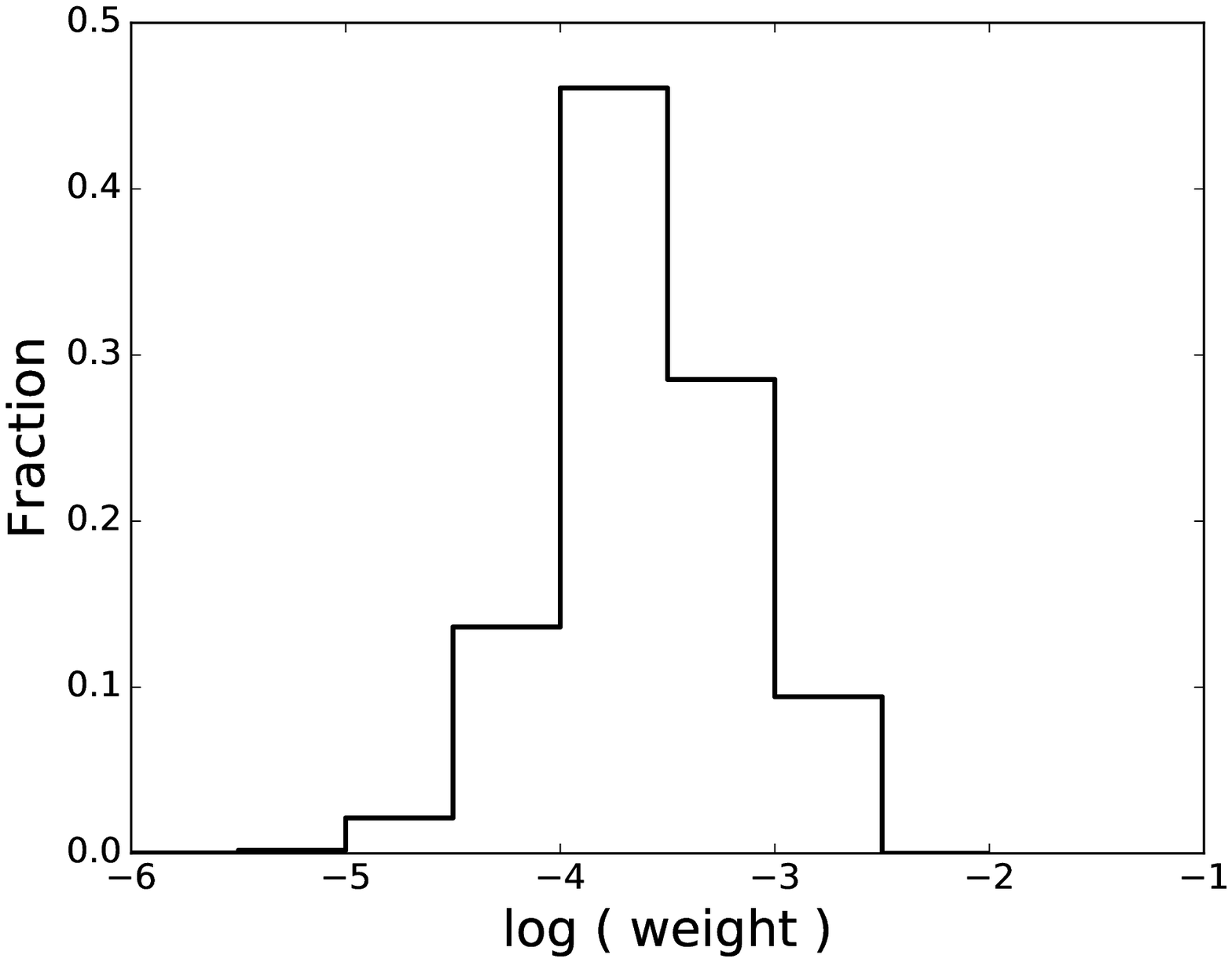} & \includegraphics[width=35mm]{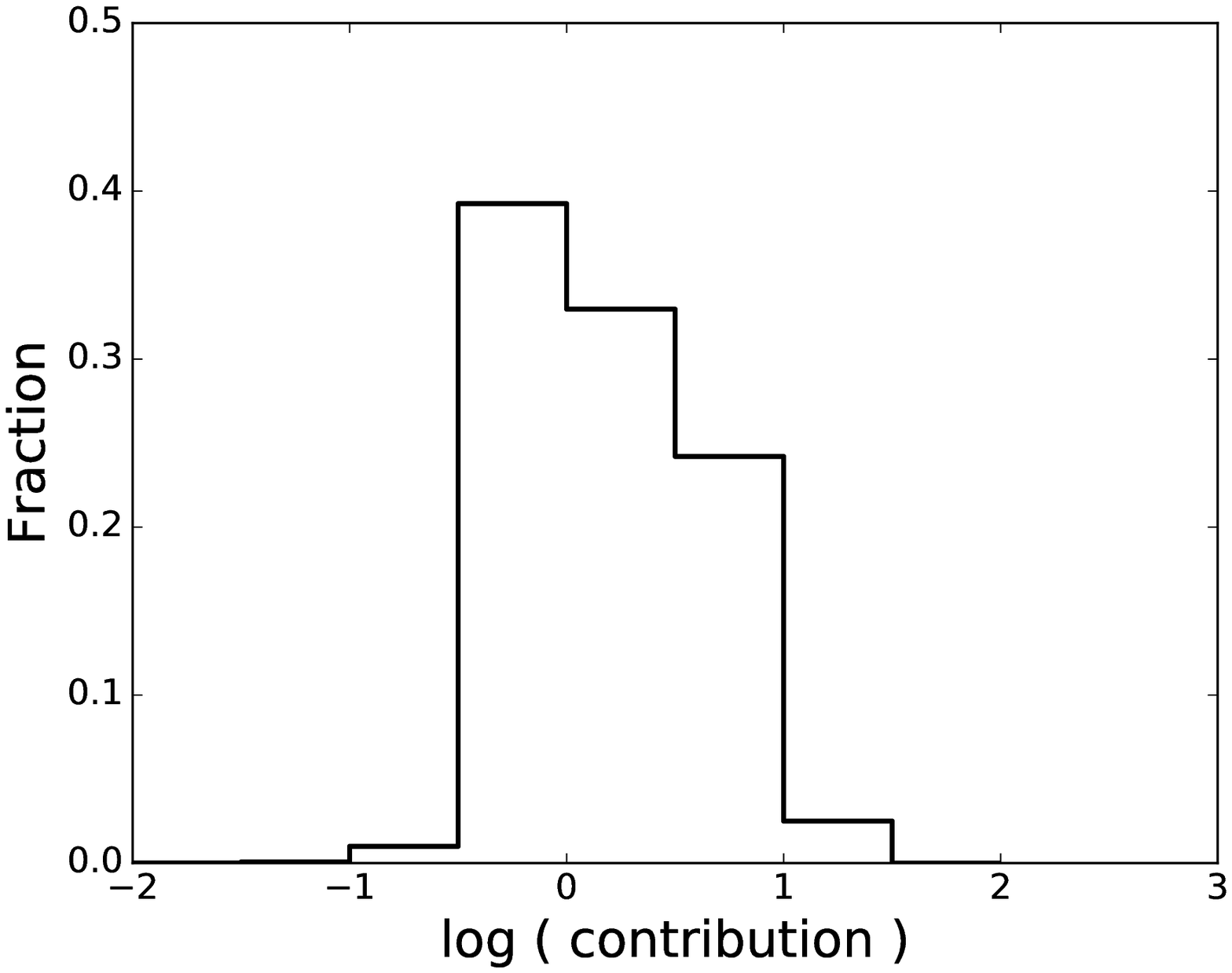} & \includegraphics[width=35mm]{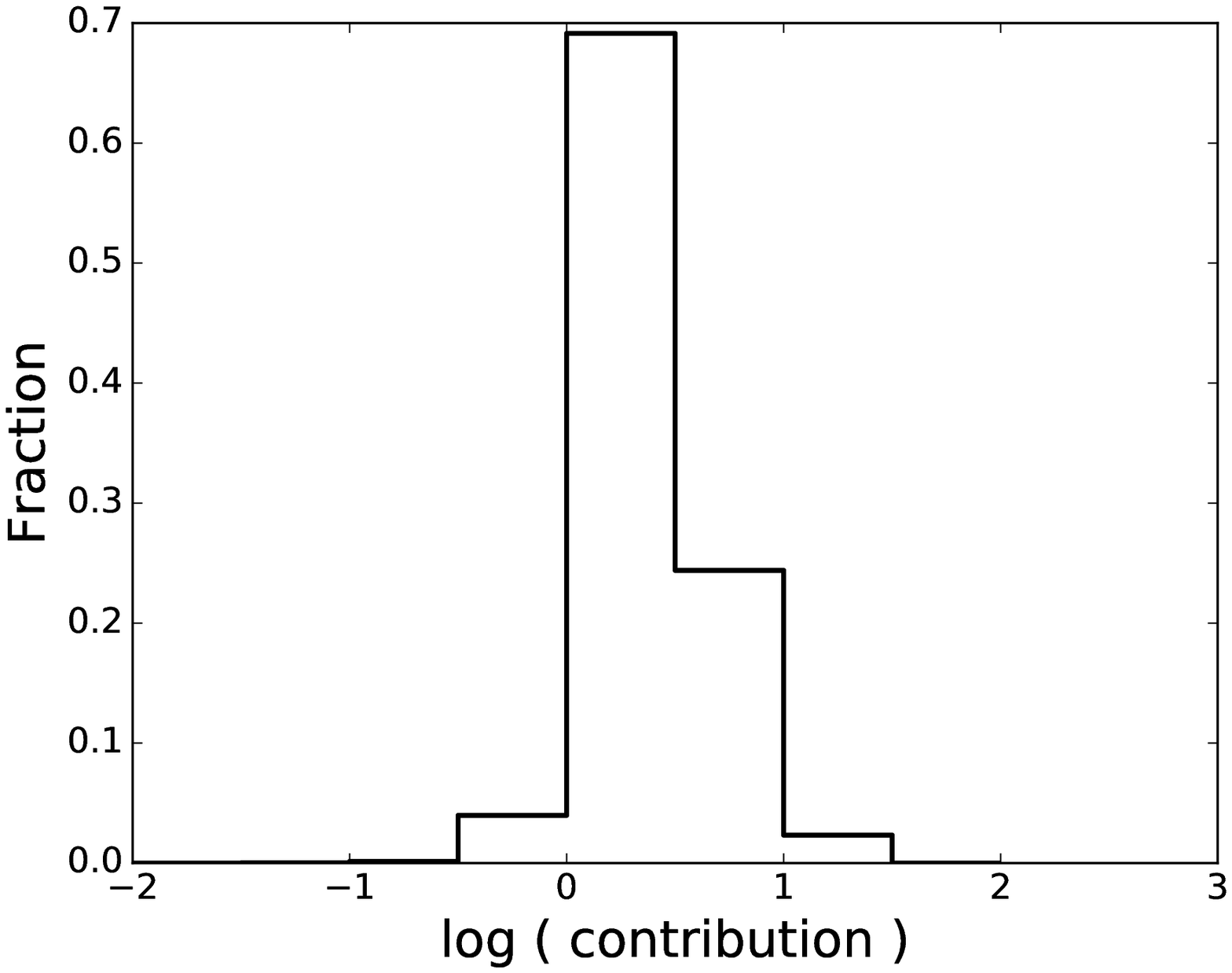} & \includegraphics[width=35mm]{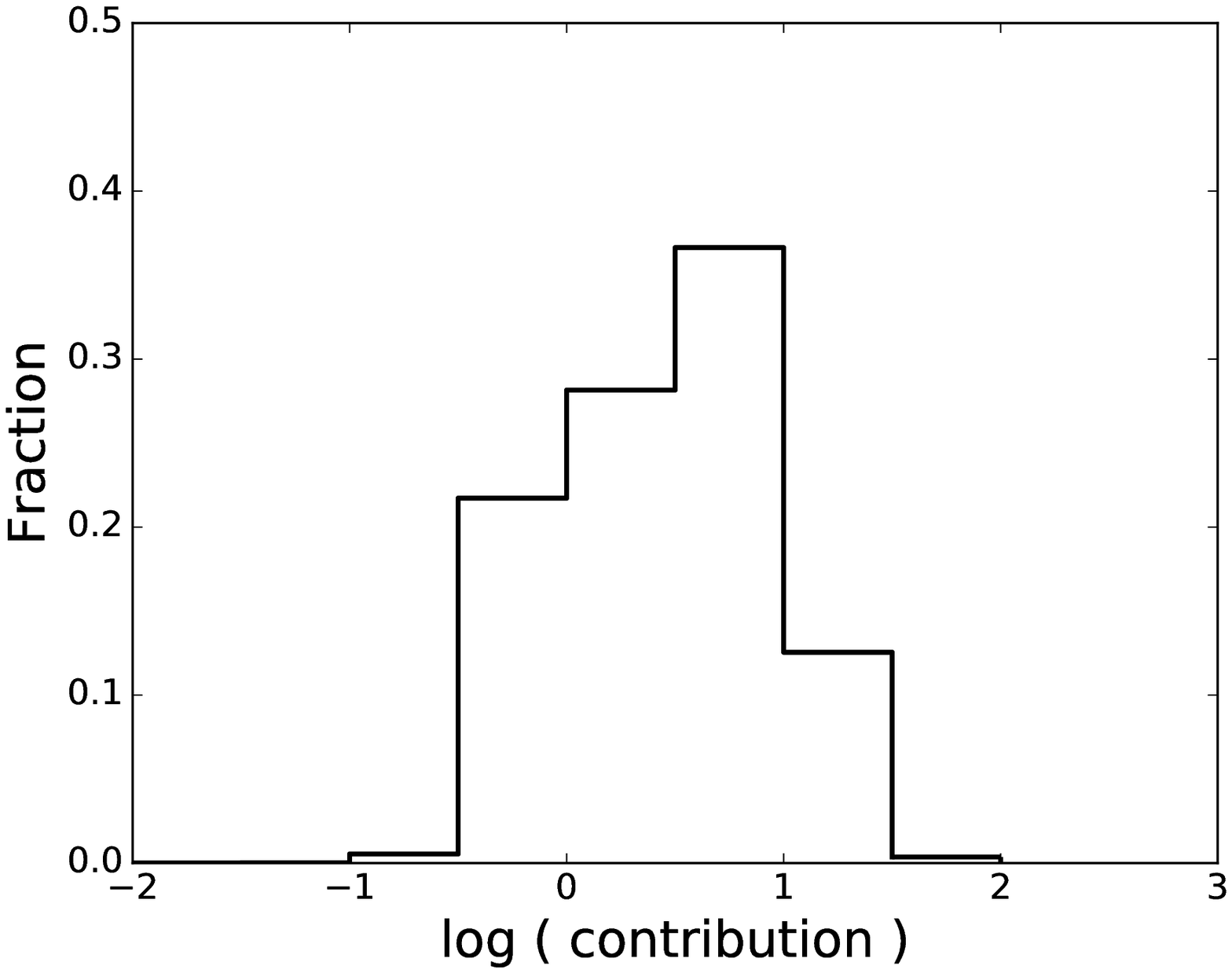} \\
\end{tabular}
\medskip
\caption{NGC 4452 orbit weight distribution and line strength contribution comparison between CVXOPT (top row) and \citet{LH1974} (bottom row). The black line indicates the fractional (weight or line strength) distribution.} 
\label{fig:NGC4452comp}
\end{figure*}

We examine the numbers of orbits contributing to the $\chi ^2$ values and show the results in Table \ref{tab:actorb}.  From the table, for \citet{LH1974}, the impact of regularisation on increasing the number of active orbits is quite clear. It is also clear that even though the line strength data is being reproduced (see Table \ref{tab:LHchi2}), not all the active orbits are contributing to all the lines.  If now we look at a comparison of the orbit weights and line strength contributions between CVXOPT and \citet{LH1974} for NGC 4452, we see that the orbit weights have the same distribution (Figure \ref{fig:NGC4452comp}).  Although the line strength profiles are similar, they are not exactly the same .  This has potential implications on just what the weighted orbits might be used for: results may be specific to the underlying techniques used (i.e. CVXOPT vs Lawson \& Hanson NNLS) and may not be totally representative of the stellar system being modelled.

\begin{table}
	\centering
	\caption{Active Orbits Comparison}
	\label{tab:actorb}
	\begin{tabular}{lcccccc}
		\hline
		       &           &  CVXOPT   & \multicolumn{4}{c}{Lawson \& Hanson NNLS} \\
		Galaxy & \# orbits & \# active & \# active& H$\beta$ & Mg$\,b$ & Fe5015 \\		
		\hline
		NGC 1248 & 7994 & 7994 & 587  & 156  & 143  & 131  \\
		         &      & 6420 & 4521 & 3383 & 2216 & 406  \\
		\hline
		NGC 3838 & 7982 & 7942 & 786  & 221  & 269  & 196  \\
		         &      & 7453 & 5824 & 4524 & 5296 & 3582 \\
		\hline
		NGC 4452 & 7997 & 7997 & 781  & 246  & 287  & 239  \\
		         &      & 7623 & 6818 & 5712 & 6252 & 6178 \\
		\hline
		NGC 4551 & 7979 & 7979 & 940  & 298  & 385  & 310  \\
		         &      & 7973 & 6496 & 4700 & 6030 & 3239 \\
		\hline	
	\end{tabular}
		
\medskip
The numbers of active orbits contributing to the orbit weight calculation and the orbit line strength contributions.  For each galaxy, the top row is without regularisation, and the bottom row with regularisation. The \# active columns gives the number of orbits with non-zero weights. The spectral line columns give the number of orbits with non-zero contributions to the model line strength values. For CVXOPT, the number of spectral line orbits is the same as the active orbits and so the CVXOPT columns are not shown.  For \citet{LH1974}, the impact of regularisation on increasing the number of active orbits is quite clear. 
\end{table}

\subsection{Data Symmetrisation}\label{sec:datasym}
Since \citet{Long2016} makes the point that chemo-M2M does not perform well with unsymmetrised data, we have evaluated how our chemo-Schwarzschild method performs.  We use symmetrised kinematic data with unsymmetrised line strength data, and our results are shown in Table \ref{tab:unsymls}.  By comparison with the symmetrised regularised models in Table \ref{tab:cvxoptchi2}, it is quite clear that unsymmetrised data does not yield acceptable models.  This may limit the applicability of our approach and is discussed further in Section \ref{sec:discuss}.
We have not yet attempted to quantify asymmetry but simple mechanisms such as axis reflection are straightforward to implement.  More sophisticated methods using, for example, the Radon transform (\citealt{Stark2018}) or symmetry pattern recognition techniques should also be considered.

\begin{table}
	\centering
	\caption{Unsymmetrised Line Strength Data - Mean $\chi^2$ Values}
	\label{tab:unsymls}
	\begin{tabular}{cccc}
		\hline
		Galaxy & H$\beta$ & Mg$\,b$ & Fe5015 \\
		\hline
		NGC 1248 & 2.51 & 1.12 & 7.47\\
		NGC 3838 & 3.05 & 1.51 & 2.57\\
		NGC 4452 & 1.76 & 1.22 & 2.77\\
		NGC 4551 & 2.03 & 1.35 & 2.40\\
		\hline
	\end{tabular}
		
\medskip
Mean $\chi^2$ values resulting from using unsymmetrised line strength data. By comparison with the regularised models in Table \ref{tab:cvxoptchi2}, the values are significantly higher ($> 1$) implying that modelling performs better with symmetrisation.
\end{table}

\section{Discussion}\label{sec:discuss}
From our results in Section \ref{sec:results}, it is clear that we are able to model successfully symmetrised 2D line strength data using our extended Schwarzschild's method. However, none of the issues identified in the chemo-M2M work in \citet{Long2016} is able to be resolved by using a chemo-Schwarzschild's method such as developed here.  We have not attempted to consider 3D aspects of modelling with Schwarzschild's method (for example, the 3D distribution of orbit line strength values): the same, previously identified concerns of uniqueness and plausability apply, and the follow up work anticipated in \citet{Long2016} concerning the likely robustness of any 3D predictions is not complete.  It now needs extending to include chemo-Schwarzschild modelling together with possible variations arising from different convex optimisation methods. In addition, the need to symmetrise data remains and cannot be resolved by the approach we have taken.  Modelling of asymmetric data remains an outstanding issue.  Perhaps chemo-Schwarzschild and chemo-M2M should only be applied to early type galaxies with thoroughly mixed stellar populations until it is addressed.

Using Schwarzschild's method as the core modelling method does bring a new set of issues as well as highlighting some additional concerns with M2M.  The lack of visibility of the impact of using the \citet{LH1974} method without regularisation is a major concern: the low number of active orbits is rarely documented in journal papers.  Based on our mean $\chi ^2$ results, Schwarzschild's method does seem to overfit the constraining observable data.  Perhaps all data (not just luminosity data) should be modelled as range constraints where the range is set using the error on the observed data. Note also that the Schwarzschild extensions do contain an implicit assumption that the orbit weights generated from modelling kinematics (with or without regularisation) are also suitable for modelling line strength data. This needs confirmation or otherwise by modelling a larger set of galaxies.

The extent to which results achieved by both the chemo-Schwarzschild and chemo-M2M approaches are influenced by the initial orbit or particle conditions requires further investigation.  Both approaches can only weight what they are provided with initially.  It is quite possible to vary the orbit mix between circular, radial and box orbits, for example, and achieve a number of plausible models reproducing the observed data. Again, more research is required.

\section{Conclusions}\label{sec:conclusions}
We have met the objectives we set out in the Introduction, Section \ref{sec:introduction}.  We have extended Schwarzschild's method into a chemo-dynamical method which is able to handle luminosity, kinematic and absorption line constraints, and have successfully applied the extended method to four ATLAS$^{\rm{3D}}$ early type galaxies.  However, as anticipated in the objectives, and as can be seen from the Discussion, Section \ref{sec:discuss}, much remains to be investigated to understand the limitations of the current approach and possible alternatives before robust predictions on real galaxies can be made.  This makes the need for follow up investigations a priority.  Overall, notwithstanding our reservations, we believe another promising step has been taken in developing a capability to perform chemo-dynamical modelling.  We now have chemo-Schwarzschild to add to our original chemo-M2M. For the future, we plan to apply these chemo-methods to additional observational data sets, such as MaNGA, to try and gain a deeper insight into the relationships between chemistry and kinematics within galaxies.

\begin{acknowledgements}
Computer runs were performed on the \textit{Venus} high performance computer system at the Tsinghua Centre for Astrophysics, Tsinghua University, China. This work is partly supported by the National Key Basic Research and Development Program of China (No. 2018YFA0404501 to SM), and by the National Science Foundation of China (Grant No. 11333003, 11390372 and 11761131004 to SM).
\end{acknowledgements}

\bibliographystyle{raa}

\bibliography{ms0238}

\end{document}